\documentclass[12pt]{article}

\usepackage{chemformula}

\usepackage{scicite}
\usepackage{amssymb}
\usepackage{graphicx}
\usepackage{subcaption}
\usepackage{url}
\usepackage{bm}
\usepackage{xcolor}
\usepackage{pdflscape}
\usepackage{times}

\newenvironment{sciabstract}{%
\begin{quote}  }
{\end{quote}}

\date{}

\begin{document} 
\begin{center}
{\LARGE A far-ultraviolet-driven photoevaporation flow}\\ 
\vspace{0.3cm}
{\LARGE  observed in a protoplanetary disk}\\[4ex]
\end{center}
Olivier Bern\'{e}$^{1,*}$,
Emilie Habart$^2$, 
Els Peeters $^{3,4,5}$,
Ilane Schroetter $^{1}$,
Am\'elie Canin $^{1}$, 
Ameek Sidhu $^{3,4}$, 
Ryan Chown $^{3,4}$, 
Emeric Bron $^{6}$,
Thomas J. Haworth $^{7}$,
Pamela Klaassen $^{8}$,
Boris Trahin $^{2}$,
Dries Van De Putte $^{9}$,
Felipe Alarc\'on $^{10}$,
Marion Zannese $^{2}$,
Alain Abergel $^{2}$,
Edwin A. Bergin $^{10}$,
Jeronimo Bernard-Salas $^{11,12}$,
Christiaan Boersma $^{13}$,
Jan Cami $^{3,4,5}$,
Sara Cuadrado $^{14}$,
Emmanuel Dartois $^{15}$,
Daniel Dicken $^{2}$,
Meriem Elyajouri $^{2}$,
Asunci\'on Fuente $^{16}$,
Javier R. Goicoechea $^{14}$,
Karl D.\ Gordon $^{9,17}$,
Lina Issa $^{1}$,
Christine Joblin $^{1}$,
Olga Kannavou $^{2}$,
Baria Khan $^{3}$,
Ozan Lacinbala $^{2}$,
David Languignon $^{6}$,
Romane Le Gal $^{1,18, 19}$,
Alexandros Maragkoudakis $^{13}$,
Raphael Meshaka $^{2}$,
Yoko Okada $^{20}$,
Takashi Onaka $^{21,22}$,
Sofia Pasquini $^{3}$,
Marc W. Pound $^{23}$,
Massimo Robberto $^{9,17}$,
Markus R\"ollig $^{20}$,
Bethany Schefter $^{3}$,
Thi\'ebaut Schirmer $^{2,24}$,
Thomas Simmer $^{2}$,
Benoit Tabone $^{2}$,
Alexander G.~G.~M. Tielens $^{23,25}$,
S\'ilvia Vicente $^{26}$,
Mark G. Wolfire $^{23}$,
\& the PDRs4All team$^{\dagger}$. 
\\
\\
\normalsize{$^{1}$Institut de Recherche en Astrophysique et Plan\'etologie, Universit\'e de Toulouse, Centre National de la Recherche Scientifique, Centre National d'Etudes Spatiales, 31028, Toulouse, France}\\
\normalsize{$^{2}$Institut d'Astrophysique Spatiale, Universit\'e Paris-Saclay, Centre National de la Recherche Scientifique, 91405 Orsay, France}\\
\normalsize{$^{3}$Department of Physics \& Astronomy, The University of Western Ontario, London ON N6A 3K7, Canada}\\
\normalsize{$^{4}$Institute for Earth and Space Exploration, The University of Western Ontario, London ON N6A 3K7, Canada}\\
\normalsize{$^{5}$Carl Sagan Center, Search for ExtraTerrestrial Intelligence  Institute, Mountain View, CA 94043, USA}\\
\normalsize{$^{6}$Laboratoire d'Etudes du Rayonnement et de la Matière, Observatoire de Paris, Université Paris Science et Lettres, Centre National de la Recherche Scientifique, Sorbonne Universit\'es, F-92190 Meudon, France}\\
\normalsize{$^{7}$Astronomy Unit, School of Physics and Astronomy, Queen Mary University of London, London E1 4NS, UK}\\
\normalsize{$^{8}$UK Astronomy Technology Centre, Royal Observatory Edinburgh, Blackford Hill EH9 3HJ, UK}\\
\normalsize{$^{9}$Space Telescope Science Institute, Baltimore, MD 21218, USA}\\
\normalsize{$^{10}$Department of Astronomy, University of Michigan, Ann Arbor, MI 48109, USA}\\
\normalsize{$^{11}${\it ACRI-ST}, Centre d’Etudes et de Recherche de Grasse, F-06130 Grasse, France}\\
\normalsize{$^{12}$Innovative Common Laboratory fo  Space Spectroscopy, 06130 Grasse, France}\\
\normalsize{$^{13}$ NASA Ames Research Center, Moffett Field, CA 94035-1000, USA}\\
\normalsize{$^{14}$Instituto de F\'{\i}sica Fundamental  (Consejo Superior de Investigacion Cientifica), 28006, Madrid, Spain}\\
\normalsize{$^{15}$Institut des Sciences Mol\'eculaires d'Orsay, Universit\'e Paris-Saclay, Centre National de la Recherche Scientifique, 91405 Orsay, France}\\
\normalsize{$^{16}$Centro de Astrobiolog\'{\i}a, Consejo Superior de Investigacion Cientifica-INTA, 28850, Torrej\'on de Ardoz, Spain}\\
\normalsize{$^{17}$Johns Hopkins University, Baltimore, MD, 21218, USA}\\
\normalsize{$^{18}$Institut de Plan\'etologie et d'Astrophysique de Grenoble, Universit\'e Grenoble Alpes, Centre National de la Recherche Scientifique, F-38000 Grenoble, France}\\
\normalsize{$^{19}$ Institut de Radioastronomie Millim\'etrique, F-38406 Saint-Martin d'H\`{e}res, France}\\
\normalsize{$^{20}$I. Physikalisches Institut, Universit\"{a}t zu K\"{o}ln, 50937 K\"{o}ln, Germany}\\
\normalsize{$^{21}$Department of Astronomy, Graduate School of Science, The University of Tokyo, Tokyo 113-0033, Japan}\\
\normalsize{$^{22}$Department of Physics, Faculty of Science and Engineering, Meisei University, Hino, Tokyo 191-8506, Japan}\\
\normalsize{$^{23}$Astronomy Department, University of Maryland, College Park, MD 20742, USA}\\
\normalsize{$^{24}$Department of Space, Earth and Environment, Chalmers University of Technology, Onsala Space Observatory, SE-439 92 Onsala, Sweden}\\
\normalsize{$^{25}$Leiden Observatory, Leiden University, P.O. Box 9513, 2300 RA Leiden, The Netherlands}\\
\normalsize{$^{26}$Instituto de Astrof\'isica e Ci\^{e}ncias do Espa\c co, P-1349-018 Lisboa, Portugal}\\
$^{\dagger}$ PDRs4All team authors and affiliations are given in the supplementary materials\\

\normalsize{$^\ast$To whom correspondence should be addressed; E-mail:  olivier.berne@irap.omp.eu}

\newpage

\baselineskip24pt

\begin{sciabstract}

Most low-mass stars form in stellar clusters that also contain massive stars, which are sources of far-ultraviolet (FUV) radiation. Theoretical models predict that this FUV radiation produces photo-dissociation regions (PDRs) on the surfaces of protoplanetary disks around low-mass stars, impacting planet formation within the disks. We report JWST and Atacama Large Millimetere Array observations of a FUV-irradiated protoplanetary disk in the Orion Nebula. Emission lines are detected from the PDR; modelling their kinematics and excitation allows us to constrain the physical conditions within the gas. We quantify the mass-loss rate induced by the FUV irradiation, finding it is sufficient to remove gas from the disk in less than a million years. This is rapid enough to affect giant planet formation in the disk.

\end{sciabstract}

Young low-mass stars are surrounded by disks of gas and dust (protoplanetary disks).
These disks have lifetimes of a few million years \cite{williams2011, haisch2001disk, andrews2020observations} and 
are the sites of planet formation \cite{keppler2018discovery}. Planet formation is
limited by processes that remove mass from the disk such as  photevaporation \cite{gorti2016disk}. 
This occurs when the upper layers of protoplanetary disks are heated by
X-ray or ultraviolet photons. 
%(of energy $E\gtrsim100$~eV), Extreme UV (EUV; 13.6~eV$< E\lesssim$100~eV), and 
%Far UV (FUV; $E< 13.6$ eV). 
Radiative heating increases the gas temperature, bringing
the local sound speed above the escape velocity of the disk, causing the gas to escape.
The photons could be from the central star \cite{gorti2008photoevaporation} or
from nearby massive stars \cite{Adams04Photo}. %, or from the interstellar radiation field \cite{Adams04Photo}.
{Because most low mass stars form in clusters that also contain massive stars, the majority 
of protoplanetary disks are exposed to radiation, so are expected to experience 
photoevaporation driven by ultraviolet photons during their lifetime \cite{lada2003, Adams04Photo, fatuzzo2008uv, winter2020prevalent, winter2022external}.
Theoretical models predict that far-ultraviolet  (FUV) photons with energies below the Lyman limit 
($E< 13.6$ eV) dominate the photoevaporation, which
affects the disk mass, radius, and lifetime \cite{storzer1999, scally2001destruction, Adams04Photo, 2007MNRAS.376.1350C, 2019MNRAS.490.5678C, 2018MNRAS.478.2700W, nicholson2019rapid, winter2020prevalent , 2022MNRAS.514.2315C},
its chemical evolution \cite{walsh2013IrrDiscChem, 2021MNRAS.503.4172H, boyden2023chemical}, and the growth and migration of any 
planet forming within the disk \cite{winter2022growth}.

However, these processes have not been directly observed. 
Most observational constraints on the mass loss rates associated have been obtained 
for ``proplyds'' in the Orion Nebula where the ionization of FUV driven photoevaporation flows 
from disks results in cometary-shaped ionization fronts (IFs) \cite{ODellWen94ONC, ODell1998Proplyds}. 
Modelling of the observed IFs has indicated mass loss rates $\dot{M}$ in units of 
solar masses (1 M$_{\odot} = 1.9891 \times 10^{30}$ kg) per year  of proplyds 
in the range $\dot{M} \approx 10^{-8}$ to $10^{-6}$ M$_{\odot} \rm{yr}^{-1}$
\cite{johnstone1998photoevaporation, henney1998modeling, Henney1999KeckONC}.
However, those observations did not determine 
the physical conditions (radiation field, gas temperature and density) 
at the location where the photoevaporation flow is launched.
In the regions where FUV photons penetrate the disk a photodissociation region (PDR) \cite{tielens1985photodissociation} 
forms at the disk surface.
Most observational tracers of PDR physics (lines of H$_2$, O and C$^{+}$) are 
in the near- and far-infrared wavelength ranges.
The spatial scale of PDRs in externally illuminated disks is 
a few hundred astronomical units ($1~\rm{au} = 1.49\times10^{6}$ meters) 
corresponding to angular sizes $<$1 arcsecond ($^{\prime\prime}$) even for the closest star forming clusters 
\cite{chen1997, storzer1999, champion2017herschel}.}

\section*{Imaging of a photoevaporation flow}

\begin{figure*}[!h]
    \centering
     \includegraphics[width=\hsize]{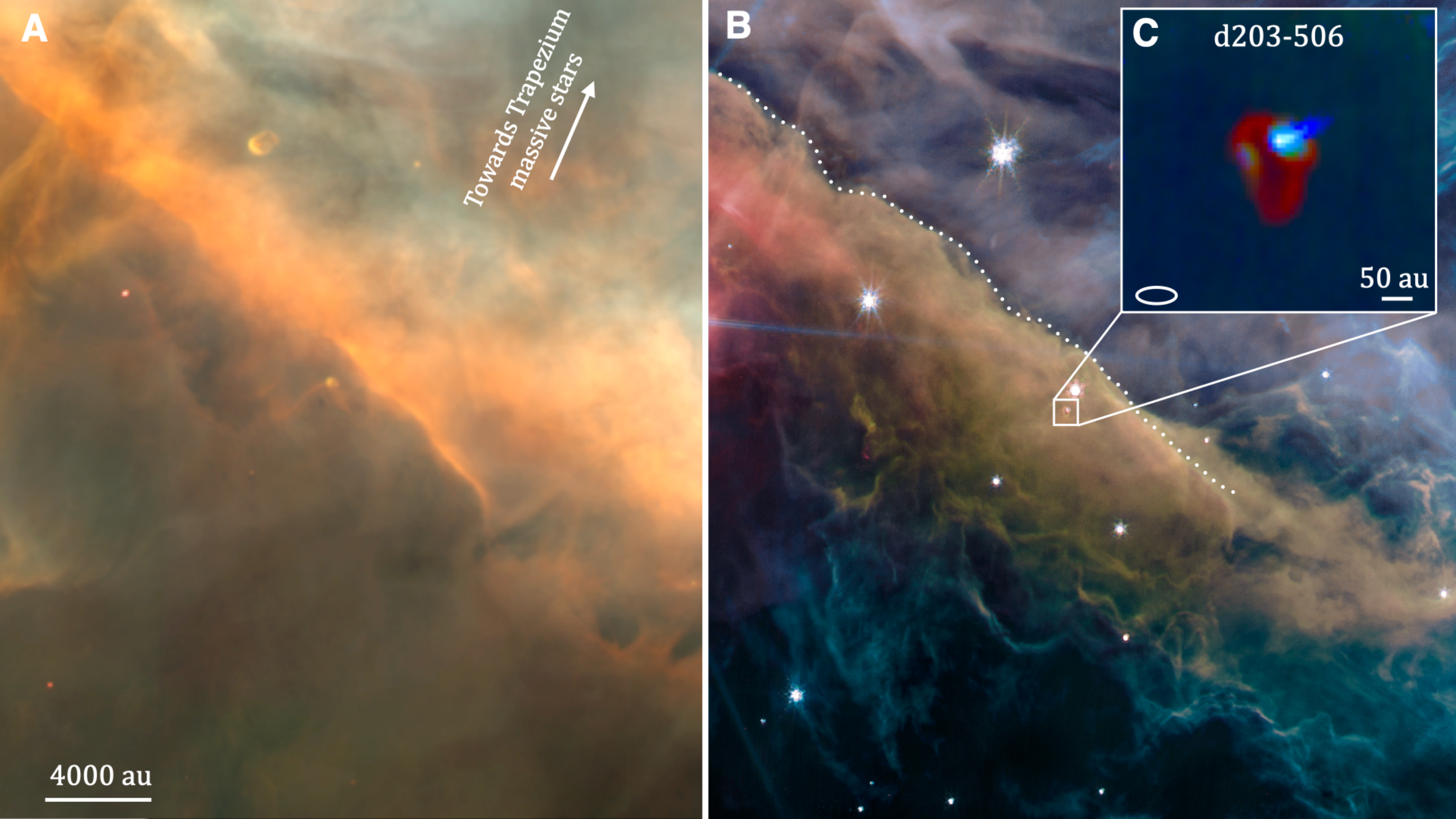}
    \caption{ {\bf Optical and near-infared images of the Orion Bar region} ({\bf A}) Hubble space telescope optical image. In  
blue is [O$\textsc{iii}$] at 502 nm, green H$\alpha$ 
at	656 nm	and red [N$\textsc{ii}$] at	658 nm.
({\bf B}) JWST near-infrared image of the same region at the same scale. Filters centered at 1.4 and 2.0 \textmu m are in blue; at 2.77, 3.00, 3.23, 3.35 and 3.32 \textmu m in green; 4.05 \textmu m in orange; and 4.44, 4.80 and 4.70 in red.
The fields of view of the images in ({\bf A}) and ({\bf B}) are centered at coordinates RA =  $5^{\rm h} 35^{\rm m} 20^{\rm s}.183$ and Dec = $-5^{\circ}25^{\prime}06^{\prime\prime}.14$. ({\bf C}) Zoom-in on the d203-506 disk. Red is the emission in the JWST-NIRCam 2.12 \textmu m  filter, tracing molecular hydrogen,  blue is the 1.64 \textmu m filter tracing the emission of [Fe$\textsc{ii}$], and green is the emission in the 1.40 \textmu m broad-band filter that traces scattered light. 
Panel ({\bf A}) credits : NASA/STScI/Rice Univ./C.O’Dell et~al \cite{image_odell}.
}
    \label{fig_overview-203-506}
\end{figure*}

Fig.~\ref{fig_overview-203-506} 
shows optical and near-infrared images of the Orion Bar,
 a ridge in the 
Orion Nebula \cite{goicoechea_compression_2016},
situated about 0.25~parsec (pc, $1~\rm{pc} = 3,086 \times 10^{16} \rm{m}$) southeast of the Trapezium Cluster of massive 
stars.  The western edge of the bar constitutes 
the ionization front (Fig.~\ref{fig_overview-203-506}~B), which separates regions where the gas is fully ionized with T$\sim$10$^{\rm 4}$~K (upper right part of the image) and the neutral atomic region with T$\sim$500-1000~K (the lower left part of the image). 
We investigate the source ``d203-506'' \cite{Bally2000ONC, berne2023formation}
with coordinates: right ascension RA =  $5^{\rm h} 35^{\rm m} 20^{\rm s}.357$  
and declination Dec = $-5^{\circ}25^{\prime}05^{\prime\prime}.81$,
a protoplanetary disk seen in absorption against the bright background. 
Previous observations of d203-506 did not show any sign 
of the presence of an ionization front \cite{Bally2000ONC, habart2022high, berne2023formation} 
indicating that the radiation field reaching the disk is  dominated by FUV photons. 

We obtained high angular resolution ($\sim 0.1^{\prime\prime}$, corresponding to $\sim 40$ au at 
the distance to Orion) images of d203-506 with the JWST and the Atacama Large Millimetere Array (ALMA). 
The JWST images were obtained in the near-infrared in multiple broad and narrow band filters using the 
near-infrared camera (NIRCam, \cite{methods}). The ALMA interferometric images provide  observations of rotational 
lines of HCN and HCO$^+$ at a velocity resolution of 0.2 km~s$^{-1}$ \cite{methods}. We also obtained spectroscopic 
observations in the near-infrared using the integral field unit (IFU) of the near-infrared Spectrograph 
(NIRSpec, \cite{methods}) on the JWST. 

\begin{figure*}[!h]
    \centering
     \includegraphics[width=17cm]{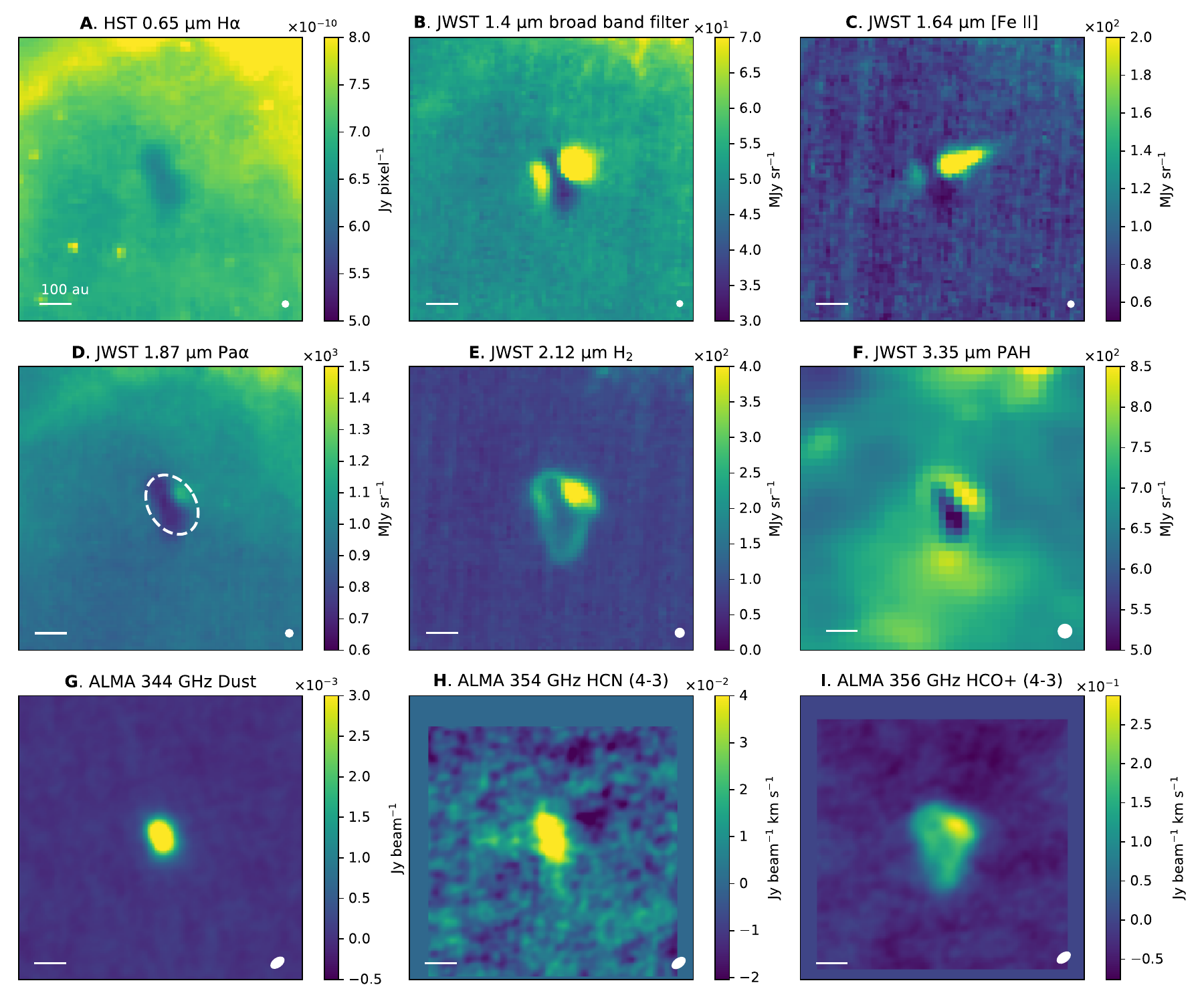}
    \caption{{ \bf Multi-wavelength images of the d203-506 disk.}  ({\bf A}) Optical image from HST \cite{ODellWen94ONC}, in a H$\alpha$ filter. ({\bf B-F}) Near-infrared images from JWST \cite{methods}. Panel {\bf E} is reproduced from \cite{berne2023formation} with permission. ({\bf G-I}) Sub-millimetre images from ALMA  \cite{methods}. In all panels the white filled ellipse indicates the size and shape of the point spread function or the reconstructed beam of the telescope and the horizontal bar is 100 au. The white dashed ellipse in panel ({\bf D}) indicates the shape of the aperture used for the extraction of the NIRSpec spectrum \cite{methods}. In panels ({\bf H}) and ({\bf I}), the notation (4-3) corresponds to the transition from quantum levels $v=0\rightarrow0, J=4\rightarrow3$. 
    The wavelength and physical assignment of each image is labelled above each panel. 1 Jy = $1\times 10^{-26} {\rm W m}^{-2}{\rm Hz}^{-1}$.}
    \label{fig_images-203-506}
\end{figure*}

\begin{figure}[!h]
    \centering
     \includegraphics[width=15cm]{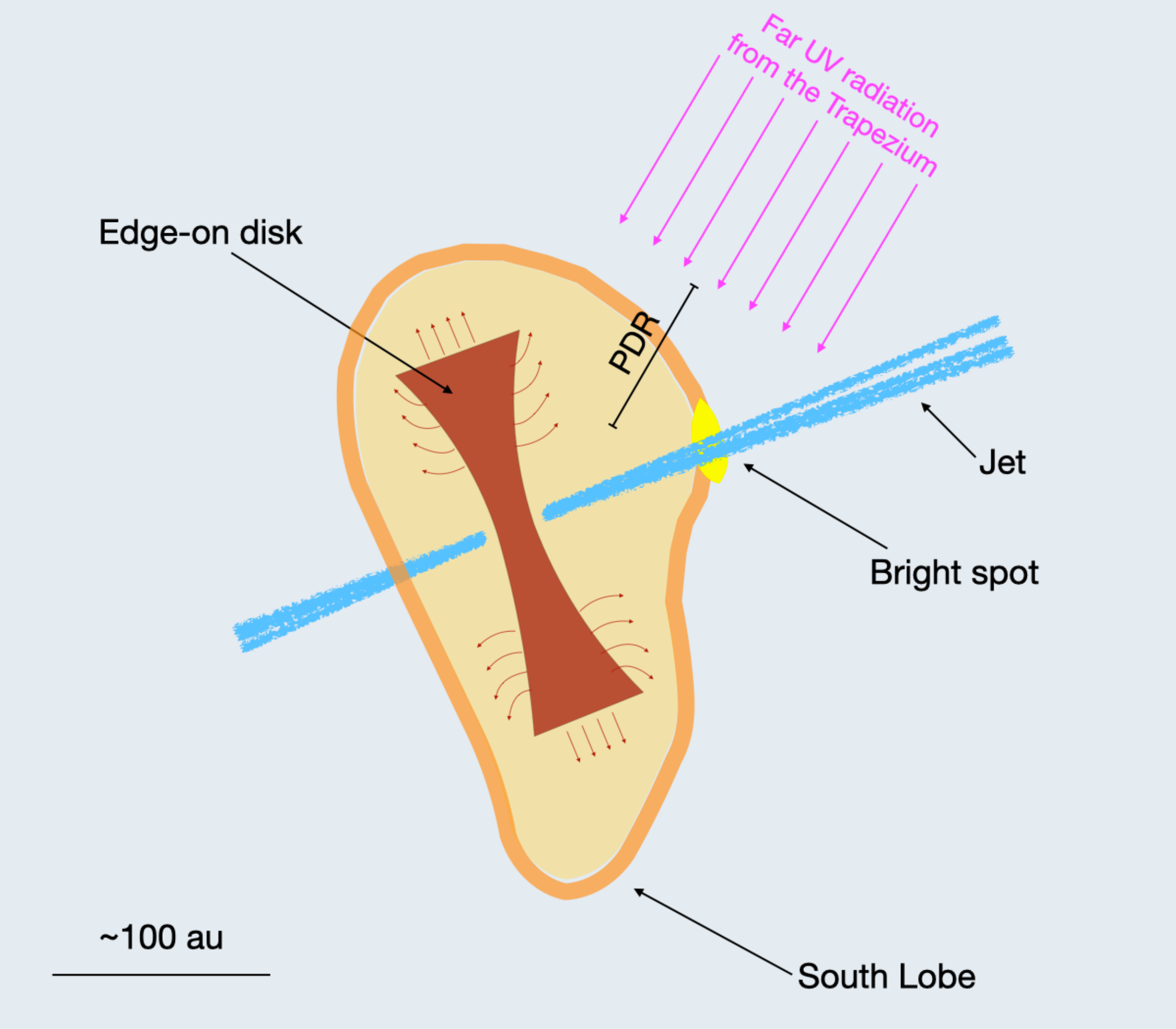}
    \caption{{\bf Schematic diagram of our interpretation of d203-506.} The edge-on disk consisting of cold molecular gas, corresponding to absorption in the NIRcam images and HCN and dust emission in ALMA images (Fig.~\ref{fig_images-203-506}), is shown in dark brown. Molecular gas escapes from this disk (brown arrows), feeding the photo-evaporation flow which creates an envelope around the disk (light brown). This envelope is delimited by the dissociation front, in orange, where molecular hydrogen is dissociated into hydrogen atoms by the far ultraviolet photons from the trapezium star (pink arrows). This transition from molecular gas of the disk to atomic gas under ultraviolet irradiation constitutes the photodissociation region (PDR). A jet from the 
    central star, shown in blue and corresponding to [Fe$\textsc{ii}$] emission interacts with the envelope creating a a bright emission spot, shown in yellow. The surroundings of d203-506 (in gray in this diagram) consist of diffuse atomic gas. }
    \label{fig_schematic}
\end{figure}

We compare the JWST and ALMA images to archival optical images from the Hubble space telescope (HST) in Fig.~\ref{fig_images-203-506}.
The nearly edge-on \cite{methods}
dusty disk is visible in absorption in all the HST and JWST images (Fig.~\ref{fig_images-203-506} { A}-{F})
but in emission in the 344 GHz (870 {\textmu}m) continuum which is due to dust 
(Fig.~\ref{fig_images-203-506}G). It is also seen in emission with ALMA in the HCN 
($v=0, J=4 \rightarrow 3$) line where $v$ and $J$ denote the 
vibrational and rotational quantum numbers, respectively, at a frequency of 354.505 GHz (or wavelength $\lambda =845.664$ {\textmu}m, Fig.~\ref{fig_images-203-506}H), 
which traces cold molecular gas.
ALMA HCO$^+$ ($v=0, J=4\rightarrow3$) at 356.734 GHz ($\lambda =840.381$ {\textmu}m)
and NIRCam H$_{\rm 2}$ ($v=1\rightarrow0, J=3\rightarrow1)$ at 2.12 {\textmu}m 
emission maps (Figs.~\ref{fig_images-203-506}I and ~\ref{fig_images-203-506}E, respectively) 
trace emission  from {  the PDR surrounding the disk} and absorption in the center. 
Both H$_{\rm 2}$ ro-vibrational and HCO$^+$ rotational emission
lines trace warm (gas kinetic temperatures $T_{\rm gas}$= 500-1000~K) molecular 
gas in PDRs \cite{goicoechea_compression_2016}.
The {PDR} is also bright in the 3.35 {\textmu}m NIRCam filter (Fig.~\ref{fig_images-203-506}F) 
dominated by aromatic infrared band (AIB) emission from ultraviolet-excited Polycyclic Aromatic Hydrocarbon (PAH) molecules. PAHs are known to be tracers of PDRs \cite{peeters_rich_2002}, 
and have been previously mapped in a proplyd in the Orion Nebula \cite{vicente2013polycyclic}. 
The {PDR in d203-506 is spatially resolved} and extends south from the disk in a lobe shape.
A jet is also clearly visible in the NIRCam [Fe~\textsc{ii}] filter  
at 1.62 {\textmu}m (Fig.~\ref{fig_images-203-506}C). 
A bright emission spot is present in the H$_{\rm 2}$ and HCO$^+$ images in the northwestern part of the PDR. 
This spot is also visible in the broad-band filter at 1.4 {\textmu}m (Fig.~\ref{fig_images-203-506}B) and appears to coincide with 
the region of interaction between the jet and the PDR, which is visible only on the side facing the Trapezium.
There is also AIB emission in the 3.35 {\textmu}m NIRCam filter at this location (Fig.~\ref{fig_images-203-506}F), 
indicating ultraviolet excitation.
Fig.~\ref{fig_schematic}  shows a schematic diagram of our interpretation of the morphological features in d203-506.

\section*{Physical properties of the PDR}
\label{sec_physical-properties}

\begin{figure}[!h]
    \centering
     \includegraphics[width=18cm]{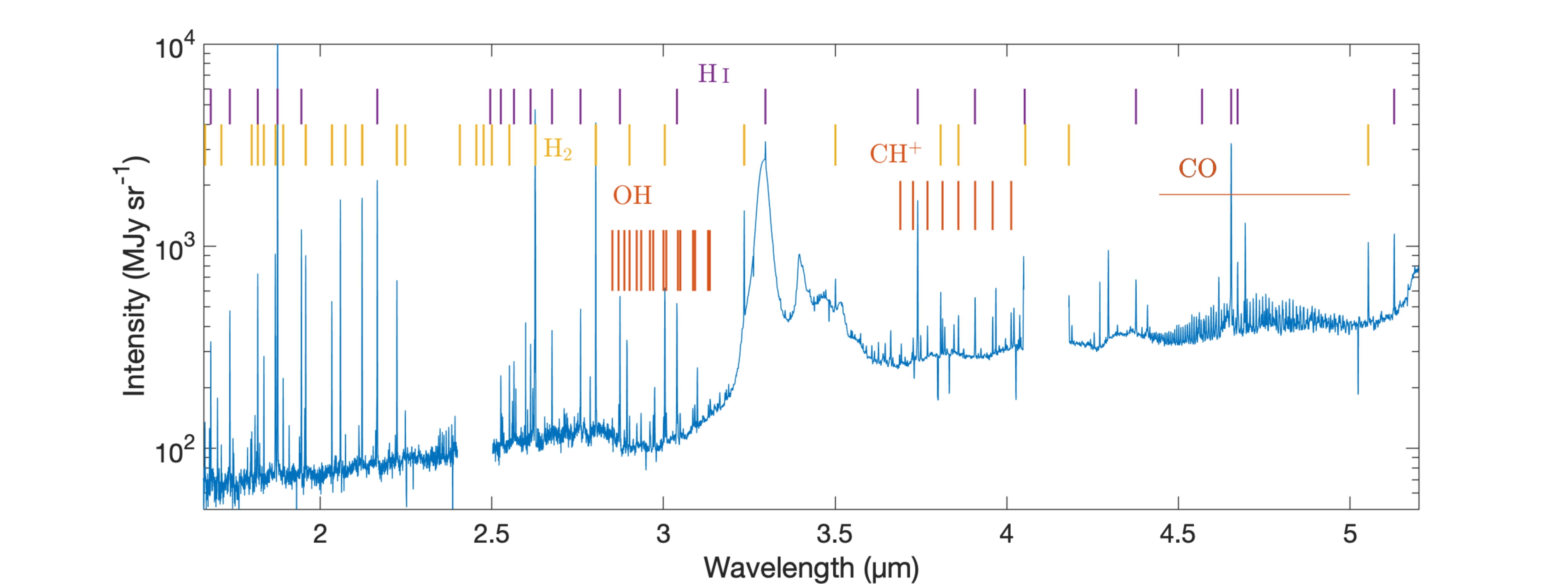}
    \caption{{ \bf JWST NIRSpec spectrum of d203-506.} Wavelength positions of the detected species are indicated. The broad emission bands at 3.3 and 3.4 {\textmu}m are from the C-H vibrational emission of polycyclic aromatic hydrocarbon (PAH) molecules. Unlabeled 
    lines are, in most cases, atomic lines (e.g., [O$\textsc{i}$] or [Fe$\textsc{ii}$]).
    There are no data between wavelengths 2.40-2.50 \textmu m and 4.05-4.18 \textmu m due to gaps in the NIRSpec detectors.}
    \label{fig_NIRSpec-spectrum}
\end{figure}

\begin{figure}[!h]
    \centering
     \includegraphics[width=15cm]{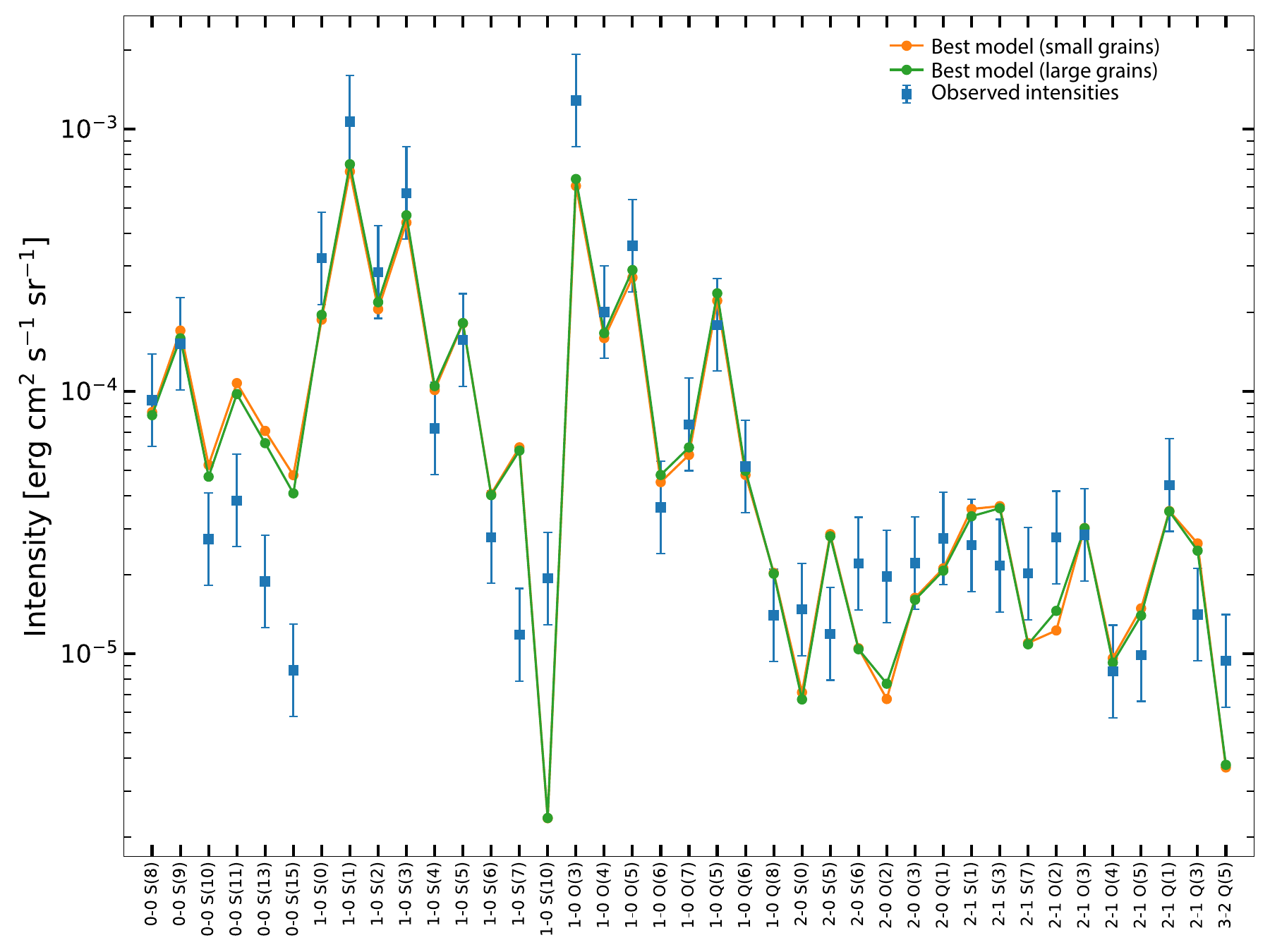}
    \caption{{ \bf Comparison of the observed and modeled H$_2$ line intensities for d203-506.} Observed intensities are the blue squares, the associated error bars represent a total  uncertainty of 50\%, as considered in the estimate of the $\chi^2$ \cite{methods}. 
    The instrumental uncertainties are smaller than the markers, and given in table~S2. Modelled intensities
    for the best models obtained using the \textsc{meudon pdr}  code are shown with the orange (for the a model using small dust grains) and green circles, for models using small and large dust grains, respectively \cite{methods}.  The notations of the H$_2$ lines in the X axis are abbreviated, quantum levels corresponding to this notation are listed in table~S2.}
    \label{fig_H2_PDR_models_intensity_diagram}
\end{figure}

Fig.~\ref{fig_NIRSpec-spectrum} shows the NIRSpec spectrum of d203-506 \cite{methods}. 
Numerous ro-vibrational emission lines of CO ($v=1\rightarrow0$ and $v=2\rightarrow1$), OH ($v=1\rightarrow0$), 
CH$^+$ ($v=1\rightarrow0$) and H$_{\rm 2}$ (up to $J=15$) are detected. 
We interpret them as coming from the PDR, so trace the physical conditions of gas in the PDR.
We model \cite{methods} the H$_{\rm 2}$ lines using the \textsc{Meudon PDR} code \cite{le_petit_model_2006}, which calculates the H$_2$ excitation in PDRs (Fig.~\ref{fig_H2_PDR_models_intensity_diagram}). 
We derive the Hydrogen number density $n_H$
and temperature of the gas in the H$_2$ emitting layer (Fig.~\ref{fig_images-203-506}E) 
as $n_H = 5.5\times10^{5} - 1.0 \times10^{7}$ and $T_{\rm gas} = 1240 - 1260$ K.
%We analyze the CH$^+$, OH and CO ro-vibrational emission using
%local thermal equilibrium radiative transfer modeling \cite{methods}
%which indicates a gas temperature that is consistent with that derived from
%the  H$_{\rm 2}$ ro-vibrational lines. 
We fitted a Keplerian model to the HCN 
emission observed with ALMA \cite{methods}. This allowed to set an upper limit 
for the mass of the central star of d203-506 $M_{\star} < 0.3 $~M$_{\odot}$ \cite{methods}. 
With $T_{\rm gas}\sim$1250~K as determined above, the speed of sound 
$c_{\rm S} \equiv \sqrt{ \frac{7/5 k_{\rm B} T_{\rm gas} }{\mu~m_{\rm H}}}$=3.3~km~s$^{-1}$,
where $k_{\rm B}$ is the Boltzmann constant, $m_{\rm H}$ is the mass of hydrogen and
$\mu$ is the ratio of total mass over hydrogen mass of interstellar gas
($\mu =1.4$ \cite{draine2011}).

This value of $c_{\rm S}$ exceeds the escape velocity at distances from the central star above 
a critical value, defined as the gravitational radius \cite{hollenbach1994}
$r_{\rm g} \equiv \frac{G~M_{\star}}{c_{\rm S}^2}$. For $M_{\star} < 0.3 $~M$_{\odot}$,
and $T_{\rm gas} = 1250~\rm{K}$, $r_{\rm g}  < 26$~au. 
This is much smaller than the observed radial
extent of the H$_2$ emission, which has a radius $r_{{\rm H}_2} = 132 \pm 13$ au
(and height $h_{{\rm H}_2} =56 \pm 13$ au, \cite{methods}).
Therefore, the gas in this layer is not gravitationally bound
and flows outwards of the disk, roughly at the speed of sound. The associated mass flux 
through the PDR is thus $j= \mu m_{\rm H} n_{\rm H} c_{\rm S}$, and the total mass loss rate
is $\dot{M} = j \times S$ where $S$ is the surface area of the H$_2$ emitting layer \cite{methods}.
Including the uncertainties on $r_{{\rm H}_2}$, $h_{{\rm H}_2} $, $n_{\rm H}$, 
and $T_{\rm gas}$ (Table S1 \cite{methods}) we calculate
$\dot{M} = 1.4 \times10^{-7}$ to $ 4.6 \times 10^{-6}$ ${\rm M}_{\odot} {\rm yr}^{-1}$.
We also investigated the mass loss rate using one-dimensional dynamical models 
finding consistent values of $\dot{M}$.

\section*{Implication for planet formation}

Gas in protoplanetary disks is the raw material from which giant planets form.
Therefore, mass-loss due to photoevaporation can limit the 
formation of such planets. 
The effects of FUV radiation depend on the stellar mass, 
which sets the strength of the gravitational field retaining the gas. 
Previous theoretical models of planet growth under the influence of external 
FUV photoevaporation predicted that  FUV radiation fields with 
intensity above about 500 times the standard interstellar radiation field 
(that is $G_{\rm 0}\gtrsim 500$ using the notation of \cite{Habing68}),
suppress giant planet formation around stars with masses 
$\lesssim$0.5~M$_{\odot}$ \cite{winter2022growth}.
Our result for d203-506 are consistent with this prediction:
it has $M_{\star} < 0.3~{\rm M}_{\odot}$,  $G_{\rm 0} \lesssim 10^{\rm 5}$ \cite{methods}
and the mass loss rate we calculated ($\dot{M} = 1.4 \times10^{-7} - 4.6 \times 10^{-6}$ M$_{\odot}$/year)
imply a disk  depletion timescale $\tau \equiv M_d / \dot{M} < 0.13$ ~Myr, with $M_d$ the disk mass \cite{methods}. 
This is faster than  even very early planet formation  \cite{stolte2015circumstellar, sheehan2018multiple}.
A positive correlation has been found between stellar mass and frequency of Jupiter-mass exoplanets \cite{johnson2010giant, reffert2015precise} which we suggest could be due to FUV radiation
in stellar clusters during the planet formation process.
Dynamical and compositional studies of Solar System bodies indicate that 
the Sun formed in a stellar cluster containing one or more 
massive stars \cite{bergin2023interstellar} so might have been affected by
FUV radiation.

\newpage 

\newpage 

\newpage 

\newpage 

\newpage

\newpage 

\bibliographystyle{Science} % style aa.bst
\bibliography{biblio_PE}

\begin{thebibliography}{10}

\bibitem{williams2011}
J.~P. Williams, L.~A. Cieza, {\it Annual Review of Astron. \& Astrophys.\/}
  {\bf 49}, 67 (2011).

\bibitem{haisch2001disk}
K.~E. Haisch~Jr, E.~A. Lada, C.~J. Lada, {\it Astrophys. J.\/} {\bf 553}, L153
  (2001).

\bibitem{andrews2020observations}
S.~M. Andrews, {\it Annual Review of Astron. \& Astrophys.\/} {\bf 58}, 483
  (2020).

\bibitem{keppler2018discovery}
M.~Keppler, {\it et~al.\/}, {\it Astron. \& Astrophys.\/} {\bf 617}, A44
  (2018).

\bibitem{gorti2016disk}
U.~Gorti, R.~Liseau, Z.~S{\'a}ndor, C.~Clarke, {\it Space Science Reviews\/}
  {\bf 205}, 125 (2016).

\bibitem{gorti2008photoevaporation}
U.~Gorti, D.~Hollenbach, {\it The Astrophysical Journal\/} {\bf 690}, 1539
  (2008).

\bibitem{Adams04Photo}
F.~C. {Adams}, D.~{Hollenbach}, G.~{Laughlin}, U.~{Gorti}, {\it Astrophys.
  J.\/} {\bf 611}, 360 (2004).

\bibitem{lada2003}
C.~J. {Lada}, E.~A. {Lada}, {\it Annual Reviews of Astron. \& Astrophys.\/}
  {\bf 41}, 57 (2003).

\bibitem{fatuzzo2008uv}
M.~Fatuzzo, F.~C. Adams, {\it Astrophys. J.\/} {\bf 675}, 1361 (2008).

\bibitem{winter2020prevalent}
A.~J. Winter, J.~D. Kruijssen, M.~Chevance, B.~W. Keller, S.~N. Longmore, {\it
  Mon. Notic. Roy. Soc.\/} {\bf 491}, 903 (2020).

\bibitem{winter2022external}
A.~J. Winter, T.~J. Haworth, {\it The European Physical Journal Plus\/} {\bf
  137}, 1132 (2022).

\bibitem{storzer1999}
H.~St{\"o}rzer, D.~Hollenbach, {\it Astrophys. J.\/} {\bf 515}, 669 (1999).

\bibitem{scally2001destruction}
A.~Scally, C.~Clarke, {\it Mon. Notic. Roy. Soc.\/} {\bf 325}, 449 (2001).

\bibitem{2007MNRAS.376.1350C}
C.~J. {Clarke}, {\it Mon. Notic. Roy. Soc.\/} {\bf 376}, 1350 (2007).

\bibitem{2019MNRAS.490.5678C}
F.~{Concha-Ram{\'\i}rez}, M.~J.~C. {Wilhelm}, S.~{Portegies Zwart}, T.~J.
  {Haworth}, {\it Mon. Notic. Roy. Soc.\/} {\bf 490}, 5678 (2019).

\bibitem{2018MNRAS.478.2700W}
A.~J. {Winter}, {\it et~al.\/}, {\it Mon. Notic. Roy. Soc.\/} {\bf 478}, 2700
  (2018).

\bibitem{nicholson2019rapid}
R.~B. Nicholson, {\it et~al.\/}, {\it Mon. Notic. Roy. Soc.\/} {\bf 485}, 4893
  (2019).

\bibitem{2022MNRAS.514.2315C}
G.~A.~L. {Coleman}, T.~J. {Haworth}, {\it Mon. Notic. Roy. Soc.\/} {\bf 514},
  2315 (2022).

\bibitem{walsh2013IrrDiscChem}
C.~{Walsh}, T.~J. {Millar}, H.~{Nomura}, {\it Astrophys. J. Let.\/} {\bf 766},
  L23 (2013).

\bibitem{2021MNRAS.503.4172H}
T.~J. {Haworth}, {\it Mon. Notic. Roy. Soc.\/} {\bf 503}, 4172 (2021).

\bibitem{boyden2023chemical}
R.~D. Boyden, J.~A. Eisner, {\it Astrophys. J.\/} {\bf 947}, 7 (2023).

\bibitem{winter2022growth}
A.~J. Winter, T.~J. Haworth, G.~A. Coleman, S.~Nayakshin, {\it Mon. Notic. Roy.
  Soc.\/}  (2022).

\bibitem{ODellWen94ONC}
C.~R. {O'Dell}, Z.~{Wen}, {\it Astrophys. J.\/} {\bf 436}, 194 (1994).

\bibitem{ODell1998Proplyds}
C.~R. {O'Dell}, {\it Astron. J.\/} {\bf 115}, 263 (1998).

\bibitem{johnstone1998photoevaporation}
D.~Johnstone, D.~Hollenbach, J.~Bally, {\it Astrophys. J.\/} {\bf 499}, 758
  (1998).

\bibitem{henney1998modeling}
W.~Henney, S.~Arthur, {\it Astron. J.\/} {\bf 116}, 322 (1998).

\bibitem{Henney1999KeckONC}
W.~J. {Henney}, C.~R. {O'Dell}, {\it Astron. J.\/} {\bf 118}, 2350 (1999).

\bibitem{tielens1985photodissociation}
A.~Tielens, D.~Hollenbach, {\it Astrophys. J.\/} {\bf 291}, 722 (1985).

\bibitem{chen1997}
H.~Chen, {\it et~al.\/}, {\it Astrophys. J.\/} {\bf 492}, L173 (1997).

\bibitem{champion2017herschel}
J.~Champion, {\it et~al.\/}, {\it Astron. \& Astrophys.\/} {\bf 604}, A69
  (2017).

\bibitem{goicoechea_compression_2016}
J.~R. Goicoechea, {\it et~al.\/}, {\it Nature\/} {\bf 537}, 207 (2016).

\bibitem{Bally2000ONC}
J.~{Bally}, C.~R. {O'Dell}, M.~J. {McCaughrean}, {\it Astron. J.\/} {\bf 119},
  2919 (2000).

\bibitem{berne2023formation}
O.~Bern{\'e}, {\it et~al.\/}, {\it Nature\/} {\bf 621}, 56 (2023).

\bibitem{habart2022high}
E.~Habart, {\it et~al.\/}, {\it arXiv preprint arXiv:2206.08245\/}  (2022).

\bibitem{methods}
 Materials and methods are available as supplementary materials.

\bibitem{peeters_rich_2002}
E.~Peeters, {\it et~al.\/}, {\it Astron. \& Astrophys.\/} {\bf 390}, 1089
  (2002).

\bibitem{vicente2013polycyclic}
S.~Vicente, {\it et~al.\/}, {\it Astrophys. J. Letters\/} {\bf 765}, L38
  (2013).

\bibitem{le_petit_model_2006}
F.~Le~Petit, C.~Nehme, J.~Le~Bourlot, E.~Roueff, {\it Astrophys. J. Suppl.
  Ser.\/} {\bf 164}, 506 (2006).

\bibitem{draine2011}
B.~T. {Draine}, {\it {Physics of the Interstellar and Intergalactic Medium}\/}
  (2011).

\bibitem{hollenbach1994}
D.~Hollenbach, D.~Johnstone, S.~Lizano, F.~Shu, {\it Astrophys. J.\/} {\bf
  428}, 654 (1994).

\bibitem{Habing68}
H.~J. {Habing}, {\it Bull. Astr. Inst. Netherlands\/} {\bf 19}, 421 (1968).

\bibitem{stolte2015circumstellar}
A.~Stolte, {\it et~al.\/}, {\it Astron. \& Astrophys.\/} {\bf 578}, A4 (2015).

\bibitem{sheehan2018multiple}
P.~D. Sheehan, J.~A. Eisner, {\it Astrophys. J.\/} {\bf 857}, 18 (2018).

\bibitem{johnson2010giant}
J.~A. Johnson, K.~M. Aller, A.~W. Howard, J.~R. Crepp, {\it Publications of the
  Astronomical Society of the Pacific\/} {\bf 122}, 905 (2010).

\bibitem{reffert2015precise}
S.~Reffert, C.~Bergmann, A.~Quirrenbach, T.~Trifonov, A.~K{\"u}nstler, {\it
  Astron. \& Astrophys.\/} {\bf 574}, A116 (2015).

\bibitem{bergin2023interstellar}
E.~A. Bergin, C.~Alexander, M.~Drozdovskaya, M.~Gounelle, S.~Pfalzner, {\it
  arXiv preprint arXiv:2301.05212\/}  (2023).

\bibitem{image_odell}
 Https://hubblesite.org/contents/media/images/2005/12/1674-Image.html.

\bibitem{zenodo_alma}
 Https://doi.org/10.5281/zenodo.8196030.

\bibitem{zenodo_nirspec}
 Https://zenodo.org/doi/10.5281/zenodo.10260214.

\bibitem{pdrcode}
 Https://zenodo.org/records/10488834.

\bibitem{casa}
 Http://casa.nrao.edu/.

\bibitem{gildas}
 Http://www.iram.fr/IRAMFR/GILDAS.

\bibitem{rieke2023performance}
M.~J. Rieke, {\it et~al.\/}, {\it Publications of the Astronomical Society of
  the Pacific\/} {\bf 135}, 028001 (2023).

\bibitem{gardner_JWST2006}
J.~P. Gardner, {\it et~al.\/}, {\it Space Science Reviews\/} {\bf 123}, 485
  (2006).

\bibitem{berne2022}
O.~{Bern{\'e}}, S.~{Foschino}, F.~{Jalabert}, C.~{Joblin}, {\it Astron. \&
  Astrophys.\/} {\bf 667}, A159 (2022).

\bibitem{jwst_pipeline}
 Https://zenodo.org/doi/10.5281/zenodo.6984365.

\bibitem{HabartE_imaging23}
E.~{Habart}, {\it et~al.\/}, {\it arXiv e-prints\/} p. arXiv:2308.16732 (2023).

\bibitem{boker2023orbit}
T.~B{\"o}ker, {\it et~al.\/}, {\it Publications of the Astronomical Society of
  the Pacific\/} {\bf 135}, 038001 (2023).

\bibitem{boker2022near}
T.~B{\"o}ker, {\it et~al.\/}, {\it Astronomy \& Astrophysics\/} {\bf 661}, A82
  (2022).

\bibitem{PeetersE_23}
E.~{Peeters}, {\it et~al.\/}, {\it arXiv e-prints\/} p. arXiv:2310.08720
  (2023).

\bibitem{menten2007distance}
K.~Menten, M.~Reid, J.~Forbrich, A.~Brunthaler, {\it Astron. \& Astrophys.\/}
  {\bf 474}, 515 (2007).

\bibitem{kurucz1992model}
R.~L. Kurucz, {\it Symposium-International Astronomical Union\/} (Cambridge
  University Press, 1992), vol. 149, pp. 225--232.

\bibitem{marconi1998near}
A.~Marconi, L.~Testi, A.~Natta, C.~Walmsley, {\it Astron. Astrophys\/} {\bf
  330}, 696 (1998).

\bibitem{boyden2020protoplanetary}
R.~D. Boyden, J.~A. Eisner, {\it Astrophys. J.\/} {\bf 894}, 74 (2020).

\bibitem{mann2014alma}
R.~K. Mann, {\it et~al.\/}, {\it Astrophys. J.\/} {\bf 784}, 82 (2014).

\bibitem{beckwith2000dust}
S.~Beckwith, T.~Henning, Y.~Nakagawa, {\it Protostars and Planets IV\/} p. 533
  (2000).

\bibitem{woitke2016}
{Woitke, P.}, {\it et~al.\/}, {\it Astron. \& Astrophys.\/} {\bf 586}, A103
  (2016).

\bibitem{andrews2005circumstellar}
S.~M. Andrews, J.~P. Williams, {\it Astrophys. J.\/} {\bf 631}, 1134 (2005).

\bibitem{roueff2019full}
E.~Roueff, {\it et~al.\/}, {\it Astron. \& Astrophys.\/} {\bf 630}, A58 (2019).

\bibitem{ismdb}
 Ismdb.obspm.fr.

\bibitem{Fitzpatrick_Massa1990}
E.~L. {Fitzpatrick}, D.~{Massa}, {\it Astrophys. J.\/} {\bf 72}, 163 (1990).

\bibitem{Wan2018}
Y.~{Wan}, {\it et~al.\/}, {\it Astrophys. J.\/} {\bf 862}, 132 (2018).

\bibitem{Bossion2018}
D.~{Bossion}, Y.~{Scribano}, F.~{Lique}, G.~{Parlant}, {\it Mon. Notic. Roy.
  Soc.\/} {\bf 480}, 3718 (2018).

\bibitem{Joblin2018}
C.~{Joblin}, {\it et~al.\/}, {\it Astron. \& Astrophys.\/} {\bf 615}, A129
  (2018).

\bibitem{Bisbas2015TORUS3DPDR}
T.~G. {Bisbas}, {\it et~al.\/}, {\it Mon. Notic. Roy. Soc.\/} {\bf 454}, 2828
  (2015).

\bibitem{2019A&C....27...63H}
T.~J. {Harries}, T.~J. {Haworth}, D.~{Acreman}, A.~{Ali}, T.~{Douglas}, {\it
  Astronomy and Computing\/} {\bf 27}, 63 (2019).

\bibitem{2013A&A...550A..36M}
D.~{McElroy}, {\it et~al.\/}, {\it Astron. \& Astrophys.\/} {\bf 550}, A36
  (2013).

\bibitem{facchini2016external}
S.~Facchini, C.~J. Clarke, T.~G. Bisbas, {\it Monthly Notices of the Royal
  Astronomical Society\/} {\bf 457}, 3593 (2016).

\end{thebibliography}

\newpage

\section*{Acknowledgments}

The author thank the anonymous referees for providing helpful feedback and comments on this paper. 
The authors thank Chris O'Dell for providing comments on this 
manuscript. 
The JWST helpdesk is acknowledged for providing support with data reduction. 
Part of this work was performed using the DiRAC Data Intensive service at Leicester, operated by the University of Leicester IT Services, which forms part of the STFC DiRAC HPC Facility.

\textbf{Funding: }

AF thanks the Spanish MICIN for funding support from PID2019-106235GB-I00. OB, IS, AC are funded by the Centre National d'Etudes Spatiales (CNES) through the APR program. JRG and SC thank the Spanish MCINN for funding support under grant PID2019-106110GB-I00. PG thanks the University Pierre and Marie Curie, the Institut Universitaire de France, the CNES, the "Programme National de Cosmologie and Galaxies" (PNCG) and the "Physique Chimie du Milieu Interstellaire" (PCMI) programs of Centre National de la Recherche Scientifique/INSU for financial supports. EP and JC acknowledge support from 
%the University of Western Ontario, 
the Institute for Earth and Space Exploration, the Canadian Space Agency, and the Natural Sciences and Engineering Research Council of Canada. 
CB is grateful for funding from
%an appointment at NASA Ames Research Center through 
the San Jos\'e State University Research Foundation (80NSSC22M0107) and acknowledges support from the Internal Scientist Funding Model (ISFM) Laboratory Astrophysics Directed Work Package at NASA Ames. TO is supported by JSPS Bilateral Program, Grant Number 120219939. TJH is funded by a Royal Society Dorothy Hodgkin Fellowship.
AF was supported by the Spanish program Unidad de Excelencia María de Maeztu CEX2020-001058-M, financed by MCIN/AEI/10.13039/501100011033. 
NN is funded through UAEU Program for Advanced Research (UPAR) grant G00003479. YZ is funded by the National Science Foundation of China (NSFC, Grant No. 11973099) and the science research grants from the China Manned Space Project (NO. CMS-CSST-2021-A09 and A10). Work by M.R. and Y.O. are supported by the Collaborative Research Centre 956, sub-project C1, funded by the Deutsche Forschungsgemeinschaft (DFG)—project ID 184018867. P.M. acknowledges grants EUR2021-122006, TED2021-129416A-I00 and PID2021-125309OA-I00 funded by MCIN/AEI/10.13039/501100011033 and European Union NextGenerationEU/PRTR. A. Pathak acknowledges financial support from the Department of Science and Technology-Core Research Grant (DST-CRG) grant (SERB-CRG/2021/000907), Institutes of Eminence (IoE) incentive grant, BHU (incentive/2021- 22/32439).
M. B. acknowledges DST INSPIRE Faculty fellowship.
%Banaras Hindu University, Varanasi and thanks the Inter-University Centre for Astronomy and Astrophysics, Pune for associateship. 
 JL is sponsored by the Chinese Academy of Sciences (CAS), through a grant to the CAS South America Center for Astronomy (CASSACA) in Santiago, Chile. HZ acknowledges support from the Swedish Research Council (contract No 2020-03437).
\\
\\
\noindent
\textbf{Author contributions: }
O. B., E. H., and E. P., led the JWST observing program. O. B. led the study and drafted the manuscript, and made Figs. 1, 3, 4, S1, S3, S4.
E. B. and F. L. P. produced the PDR models and Figs 5, S4, S5, S6, and 4. T. J. H. produced the 1d dynamical models and Fig. S7. F. A. produced the the disk 3D radiative transfer models and Fig. S2. A. Canin produced Fig. 2. 
Jason C. led PI the ALMA observing program. Jason C. and Edwige C. conducted the ALMA data reduction. P. K. anlayzed the ALMA data.
I.S., A.S., R.C., D.V.P. and F.A. reduced the NIRSpec data. 
A. C. and B. Tr. reduced the NIRCam data. 
I. S. produced the extraction of the NIRSpec spectrum.
I. S. and M. Z. extracted the line intensities. 
M. Z. provided conducted the LTE radiative transfer models for 
CH$^+$ and OH. J. C. provided the radiative transfer models for CO.
All other authors contributed to the research presented in this paper or provided detailed feedback on the manuscript. All authors meet the journal's authorship requirement.

\noindent
\textbf{Competing interests: }The authors declare no competing interests.

\noindent
\textbf{Data and material availability: }The JWST data are available on the MAST portal (\url{http://mast.stsci.edu}) using the program ID 1288. The Hubble Space Telescope data are available through the same portal, using program ID 6603. The ALMA raw data is available on the ALMA archive (\url{https://almascience.eso.org/aq/}) using program ID 2017.1.01478.S. Our reduced ALMA data are available on Zenodo
\cite{zenodo_alma}, as well as the reduced NIRSPec spectrum of d203-506 \cite{zenodo_nirspec}. 
The \textsc{torus-3dpdr} code is available at \url{https://bitbucket.org/tjharries/torus/src/master/}. The $\textsc{meudon pdr}$ code used in this paper \cite{pdrcode} is available at \url{https://zenodo.org/records/10488834}.\\

\noindent
\textbf{Supplementary materials: }\\
PDRs4All Team authors and affiliations\\
Materials and Methods\\
%Supplementary Text\\
Figures S1 to S7\\
Tables S1 to S2\\
References\\

\newpage

{\Large Supplementary Materials for} \\[4ex]
\textbf{\large A far-ultraviolet-driven photoevaporation flow observed in a protoplanetary disk}\\[4ex]
{O. Bern\'e, \textit{et al.}}\\[4ex]
{Corresponding author 
Email: olivier.berne@irap.omp.eu}
%\end{center} 
\hspace*{0.5cm}\\[4ex]
\thispagestyle{empty}
\noindent {  This PDF file includes:}\\
PDRs4All Team authors and affiliations\\
Materials and Methods\\
%Supplementary Text\\
Figures S1 to S7\\
Tables S1 to S2\\
References 50 - 78\\

\newpage
\setcounter{page}{1}
\section*{PDRs4All Team authors and affiliations}
\label{pdrs4all_team}

Olivier Bern\'{e} $^{1,*}$,
Emilie Habart $^2$, 
Els Peeters $^{3,4,5}$,
Ilane Schroetter $^{1}$,
Am\'elie Canin $^{1}$, 
Ameek Sidhu $^{3,4}$, 
Ryan Chown $^{3,4}$, 
Emeric Bron $^{6}$,
Thomas J. Haworth $^{7}$,
Pamela Klaassen $^{8}$,
Boris Trahin $^{2}$
% ...
\\
\\
\normalsize{$^{1}$Institut de Recherche en Astrophysique et Plan\'etologie, Universit\'e de Toulouse, Centre National de la Recherche Scientifique, Centre National d'Etudes Spatiales, 31028, Toulouse, France}\\
\normalsize{$^{2}$Institut d'Astrophysique Spatiale, Universit\'e Paris-Saclay, Centre National de la Recherche Scientifique, 91405 Orsay, France}\\
\normalsize{$^{3}$Department of Physics \& Astronomy, The University of Western Ontario, London ON N6A 3K7, Canada}\\
\normalsize{$^{4}$Institute for Earth and Space Exploration, The University of Western Ontario, London ON N6A 3K7, Canada}\\
\normalsize{$^{5}$Carl Sagan Center, Search for ExtraTerrestrial Intelligence  Institute, Mountain View, CA 94043, USA}\\
\normalsize{$^{6}$Laboratoire d'Etudes du Rayonnement et de la Matière, Observatoire de Paris, Université Paris Science et Lettres, Centre National de la Recherche Scientifique, Sorbonne Universit\'es, F-92190 Meudon, France}\\
\normalsize{$^{7}$Astronomy Unit, School of Physics and Astronomy, Queen Mary University of London, London E1 4NS, UK}\\
\normalsize{$^{8}$UK Astronomy Technology Centre, Royal Observatory Edinburgh, Blackford Hill EH9 3HJ, UK}\\
\normalsize{$^{9}$Space Telescope Science Institute, Baltimore, MD 21218, USA}\\
\normalsize{$^{10}$Department of Astronomy, University of Michigan, Ann Arbor, MI 48109, USA}\\
\normalsize{$^{11}${\it ACRI-ST}, Centre d’Etudes et de Recherche de Grasse, F-06130 Grasse, France}\\
\normalsize{$^{12}$Innovative Common Laboratory fo  Space Spectroscopy, 06130 Grasse, France}\\
\normalsize{$^{13}$ NASA Ames Research Center, Moffett Field, CA 94035-1000, USA}\\
\normalsize{$^{14}$Instituto de F\'{\i}sica Fundamental, Consejo Superior de Investigacion Cientifica, 28006, Madrid, Spain}\\
\normalsize{$^{15}$Institut des Sciences Mol\'eculaires d'Orsay, Universit\'e Paris-Saclay, Centre National de la Recherche Scientifique, 91405 Orsay, France}\\
\normalsize{$^{16}$Centro de Astrobiolog\'{\i}a, Consejo Superior de Investigacion Cientifica, Instituto Nacional de Técnica Aerospacial, 28850, Torrej\'on de Ardoz, Spain}\\
\normalsize{$^{17}$Johns Hopkins University, Baltimore, MD, 21218, USA}\\
\normalsize{$^{18}$Institut de Plan\'etologie et d'Astrophysique de Grenoble, Universit\'e Grenoble Alpes, Centre National de la Recherche Scientifique, F-38000 Grenoble, France}\\
\normalsize{$^{19}$ Institut de Radioastronomie Millim\'etrique, F-38406 Saint-Martin d'H\`{e}res, France}\\
\normalsize{$^{20}$I. Physikalisches Institut Universit\"{a}t zu K\"{o}ln, 50937 K\"{o}ln, Germany}\\
\normalsize{$^{21}$Department of Astronomy, Graduate School of Science, The University of Tokyo, Tokyo 113-0033, Japan}\\
\normalsize{$^{22}$Department of Physics, Faculty of Science and Engineering, Meisei University, Hino, Tokyo 191-8506, Japan}\\
\normalsize{$^{23}$Astronomy Department, University of Maryland, College Park, MD 20742, USA}\\
\normalsize{$^{24}$Department of Space, Earth and Environment, Chalmers University of Technology, Onsala Space Observatory, SE-439 92 Onsala, Sweden}\\
\normalsize{$^{25}$Leiden Observatory, Leiden University, 2300 RA Leiden, The Netherlands}\\
\normalsize{$^{26}$Instituto de Astrof\'isica e Ci\^{e}ncias do Espa\c co, P-1349-018 Lisboa, Portugal}\\
\normalsize{$^{27}$Instituto de Física e Química, Universidade Federal de Itajubá, Itajubá, Brazil}\\
\normalsize{$^{28}$Bay Area Environmental Research Institute, Moffett Field, CA 94035, USA}\\
\normalsize{$^{29}$Australian Synchrotron, Australian Nuclear Science and Technology Organisation, Victoria, Australia}\\
\normalsize{$^{30}$INAF - Osservatorio Astrofisico di Catania, 95123 Catania, Italy}\\
\normalsize{$^{31}$Department of Physics, Faculty of Science, University of Zagreb, 10000 Zagreb, Croatia}\\
\normalsize{$^{32}$Department of Space, Earth, and Environment, Chalmers University of Technology, Onsala Space Observatory, 43992, Onsala, Sweden}\\
\normalsize{$^{33}$Laboratoire de Physique de l'\'Ecole Normale Sup\'erieure, Universit\'e Paris Science et Lettres, Centre National de la Recherche Scientifique, Sorbonne Universit\'e, Universit\'e de Paris, 75005, Paris, France}\\
\normalsize{$^{34}$Laboratory for Atmospheric and Space Physics, University of Colorado, Boulder, CO 80303, USA}\\
\normalsize{$^{35}$Department of Chemistry, University of Colorado, Boulder, CO 80309, USA}\\
\normalsize{$^{36}$Institute for Modeling Plasma, Atmospheres, and Cosmic Dust, University of Colorado, Boulder, CO 80303, USA}\\
\normalsize{$^{37}$Faculty of Aerospace Engineering, Delft University of Technology, 2629 HS Delft, The Netherlands}\\
\normalsize{$^{38}$Radboud University, Institute for Molecules and Materials, Free-Electron Lasers for Infrared eXperiments Laboratory, 6525 ED Nijmegen, the Netherlands}\\
\normalsize{$^{39}$} School of Physics, University of Hyderabad, Hyderabad 500 046, India \\
\normalsize{$^{40}$Department of Physics, Wellesley College, 106 Central Street, Wellesley, MA 02481, USA}\\
\normalsize{$^{41}$Anton Pannekoek Institute for Astronomy, University of Amsterdam, Science Park 904, 1098 XH Amsterdam, The Netherlands}\\
\normalsize{$^{42}$Laboratoire de Physique des deux infinis Ir\`ene Joliot-Curie, Universit\'e Paris-Saclay, Centre National de la Recherche Scientifique, 91405 Orsay Cedex, France}\\
\normalsize{$^{43}$Institut de Physique de Rennes, Centre National de la Recherche Scientifique 6251, Universit{\'e} de Rennes 1, 35042 Rennes Cedex, France}\\
\normalsize{$^{44}$Department of Chemistry, Gandhi Institute of Technology and Management, Bangalore, India} \\
\normalsize{$^{45}$Department of Chemistry, The University of British Columbia, Vancouver, British Columbia, Canada}\\
\normalsize{$^{46}$National Radio Astronomy Observatory, 520 Edgemont Road, Charlottesville, VA 22903, USA}\\
\normalsize{$^{47}$European Space Agency, Villanueva de la Ca\~nada, E-28692 Madrid, Spain}\\
\normalsize{$^{48}$Observatoire de Paris, Paris Science et Lettres University, Sorbonne Universit\'e, Laboratoire d'Etudes du Rayonnement et de la Matière, 75014, Paris, France}\\
\normalsize{$^{49}$Harvard-Smithsonian Center for Astrophysics, 60 Garden Street, Cambridge MA 02138, USA}\\
\normalsize{$^{50}$Sorbonne Universit\'{e}, Centre National de la Recherche Scientifique, Institut d'Astrophysique de Paris,75014 Paris, France}\\
\normalsize{$^{51}$Institut Universitaire de France, Minist{\`e}re de l'Enseignement Sup{\'e}rieur et de la Recherche, 1 rue Descartes, 75231 Paris, France}\\
\normalsize{$^{52}$Department of Physics and Astronomy, Rice University, Houston TX, 77005-1892, USA}\\
\normalsize{$^{53}$Yunnan Observatories, Chinese Academy of Sciences, 396 Yangfangwang, Guandu District, Kunming, 650216, China}\\
\normalsize{$^{54}$Chinese Academy of Sciences South America Center for Astronomy, National Astronomical Observatories, Chinese Academy of Science, Beijing 100101, China}\\
\normalsize{$^{55}$Departamento de Astronom\'ia, Universidad de Chile, Santiago, Chile}\\
\normalsize{$^{56}$Departments of Chemistry and Astronomy, University of Virginia, Charlottesville, Virginia 22904, USA}\\
\normalsize{$^{57}$Center for Interstellar Catalysis, Department of Physics and Astrononmy, Aarhus University, 8000 Aarhus C, Denmark}\\
\normalsize{$^{58}$Laboratory Astrophysics Group of the Max Planck Institute for Astronomy at the Friedrich Schiller University Jena, Institute of Solid State Physics, 07743 Jena, Germany}\\
\normalsize{$^{59}$Instituto de Astronomia, Geof\'isica e Ci\^encias Atmosf\'ericas, Universidade de S\~ao Paulo, 05509-090 S\~ao Paulo, Brazil}\\
\normalsize{$^{60}$Department of Physics and Astronomy, San Jos\'e State University, San Jose, CA 95192, USA}\\
\normalsize{$^{61}$Institut de Ciencies de l'Espai, Consejo Superior de Investigacion Cientifica, E-08193, Barcelona, Spain}\\
\normalsize{$^{62}$Institución Catalana de Investigación y Estudios Avanzados, Pg. Lluís Companys 23, E-08010 Barcelona, Spain}\\
\normalsize{$^{63}$Institut d'Estudis Espacials de Catalunya, E-08034 Barcelona, Spain} \\
\normalsize{$^{64}$European Space Agency, Space Telescope Science Institute, Baltimore MD 21218, USA}\\
\normalsize{$^{65}$Institute of Astronomy, Russian Academy of Sciences, 119017, Moscow, Russia}\\
\normalsize{$^{66}$Department of Earth, Ocean, \& Atmospheric Sciences, University of British Columbia, Canada V6T 1Z4}\\
\normalsize{$^{67}$ {\it Telespazio UK}, European Space Agency, E-28692 Villanueva de la Ca\~nada, Madrid, Spain}\\
\normalsize{$^{68}$Infrared Processing and Analysis Center, California Institute of Technology,  Pasadena, CA 91125, USA}\\
\normalsize{$^{69}$Department of Physics and Astronomy, University of Missouri, Columbia, MO 65211, USA}\\
\normalsize{$^{70}$Max Planck Institute for Astronomy, K\"onigstuhl 17, 69117 Heidelberg, Germany}\\
\normalsize{$^{71}$Chemical Sciences Division, Lawrence Berkeley National Laboratory, Berkeley, California, USA}\\
\normalsize{$^{72}$Kenneth S.~Pitzer Center for Theoretical Chemistry, Department of Chemistry, University of California -- Berkeley, Berkeley, California, USA}\\
\normalsize{$^{73}$AIM, Commissariat à l'Énergie Atomique et aux Énergies Alternatives, Centre National de la Recherche Scientifique, Universit\'e Paris-Saclay, Universit\'e Paris Diderot, Sorbonne Paris Cit\'e, 91191 Gif-sur-Yvette, France}\\
\normalsize{$^{74}$Institut des Sciences Moléculaires, Centre National de la Recherche Scientifique, Université de Bordeaux, 33405 Talence, France}\\
\normalsize{$^{75}$Department of Chemistry, Massachusetts Institute of Technology, Cambridge, MA 02139, USA}\\
\normalsize{$^{76}$National Radio Astronomy Observatory, Charlottesville, VA 22903, USA}\\
\normalsize{$^{77}$Instituto de Ciencia de Materiales de Madrid, Consejo Superior de Investigacion Cientifica, E28049, Madrid, Spain}\\
\normalsize{$^{78}$Department of Physics, PO Box 64, 00014 University of Helsinki, Finland}\\
\normalsize{$^{79}$Steward Observatory, University of Arizona, Tucson, AZ 85721-0065, USA}\\
\normalsize{$^{80}$ Osservatorio Astronomico di Cagliari, Instituto Nazionale di Astrofisca, 09047 Selargius, Italy}\\
\normalsize{$^{81}$Department of Physics, College of Science, United Arab Emirates University, Al-Ain, 15551, UAE}\\
\normalsize{$^{82}$National Astronomical Observatory of Japan, Tokyo 181-8588, Japan}\\
\normalsize{$^{83}$Department of Physics, Institute of Science, Banaras Hindu University, Varanasi 221005, India}\\
\normalsize{$^{84}$Department of Physics, University of Central Florida, Orlando, FL 32816-2385, USA}\\
\normalsize{$^{85}$Van’t Hoff Institute for Molecular Sciences, University of Amsterdam, PO Box 94157, 1090 GD, Amsterdam, The Netherlands} \\
\normalsize{$^{86}$Laboratoire de Chimie et Physique Quantiques, Universit\'e de Toulouse, Centre National de la Recherche Scientifique, Toulouse, France}\\
\normalsize{$^{87}$Instituto de Matem\'atica, Estat\'istica e F\'isica, Universidade Federal do Rio Grande, 96201-900, Rio Grande, Brazil}\\
\normalsize{$^{88}$Center for Astrophysics and Space Sciences, Department of Physics, University of California, San Diego, CA 92093, USA}\\
\normalsize{$^{89}$School of Chemistry, The University of Nottingham, University Park, Nottingham, NG7 2RD, United Kingdom}\\
\normalsize{$^{90}$Astronomy Department, Ohio State University, Columbus, OH 43210 USA}\\
\normalsize{$^{91}$Space Science Institute, 4765 Walnut St., R203, Boulder, CO 80301, USA}\\
\normalsize{$^{92}$Department of Physics, Stockholm University, SE-10691 Stockholm, Sweden}\\
\normalsize{$^{93}$Department of Physics, Texas State University, San Marcos, TX 78666 USA}\\
\normalsize{$^{94}$Ritter Astrophysical Research Center, University of Toledo, Toledo, OH 43606, USA}\\
\normalsize{$^{95}$School of Physics and Astronomy, Sun Yat-sen University, Zhuhai 519000, China}\\
\normalsize{$^{96}$Star and Planet Formation Laboratory, Rikagaku Kenkyusho (RIKEN) Cluster for Pioneering Research, Saitama 351-0198, Japan}\\
\normalsize{$^{97}$University of Science and Technology of China, Chinese Academy of Science Key Laboratory of Crust-Mantle Materials and Environment, Anhui 230026, China}\\

\newpage

\section*{Materials and methods}
\label{methods}

\subsection*{Observations and data reduction}
\label{sec_observation-data-reduction}

\subsubsection*{ALMA} 
ALMA observations were taken in December 2017 as part of 
project 2017.1.01478.S (Principal Investigator J.~Champion). 
The configuration for the observations was C43-6 with 46 antennas, corresponding to 1035 
baselines ranging from 15.1~m to 3.3~km.
We observed the HCO$^+$ ($v=0, J=4\rightarrow3$) line at 356.734242~GHz and 
the HCN ($v=0, J=4\rightarrow3$) line at 354.505473~GHz, at a velocity resolution of
0.21~km~s$^{-1}$. In addition, we obtained dust continuum emission at 344.0~GHz.
{  The weather conditions were average. Low altitude antennas 
%were not subjected to high wind 
did experience humidity while observing.
The data were reduced and calibrated  
using the \textsc{casa} (version 5.5.5-5) software \cite{casa}.
%with the help of the ALMA regional node in Grenoble, France. 
For each spectral window (corresponding to observation of the continuum, HCO$^+$ and HCN), 
the measurement of system noise temperature was performed in a spectral window in time division mode 
(TDM) with low-spectral resolution: 128 channels in a bandwidth of 2 GHz. The continuum was calibrated 
in its own spectral window but the windows for the  lines were calibrated with others dedicated windows. 
We performed automatic flagging to remove outliers, incomplete data, 
or artificial lines due to systematic noise. During this procedure, we  found that antenna 
DV07 measured only noise so we discarded its data. 
The phase calibration was done by measuring the amount of water towards quasars 
QSO B0539-057 and QSO B0507+179
using a sensor of the \ch{H2O} 183GHz line in each antenna.
The calibrator for absolute flux was the quasar QSO B0507+179. 
Our observation was on December 10 2017 and the quasar was observed the day before (9 December 2017) 
with a flux value of 1.25 ± 0.08 Jy at 343.5 GHz. 
Following the calibration, the images were reconstructed. % thanks to support of the ALMA regional node in Grenoble. 
We used  the \textsc{gildas} (Oct. 2019 version) software \cite{gildas} to produce the Fourier space visibility tables, derive clean maps and subtract the local continuum around lines.  
The final beam size in these maps is 0.13$^{\prime\prime}$$\times$0.08$^{\prime\prime}$ with a position angle of 123$^{\circ}$ with respect to the East-West axis.}

\subsubsection*{JWST NIRCam}

Observations with NIRCam \cite{rieke2023performance} on the JWST \cite{gardner_JWST2006}
were obtained as part of the PDRs4All early release science (ERS) program 
\cite{berne2022} Sep. 10, 2022. 
The filters used in this paper are F140M, F164N, F187N, F121N, and F335M. 
We used the RAPID readout with 2 groups per integration, 2 integrations, and 4 dithers,
providing a total on-source exposure time of 214.73s in each filter.
These observations were reduced using the JWST pipeline version 1.7.1 \cite{jwst_pipeline} 
with calibration reference data system (CRDS) context file jwst\_0969.pmap. 
No sky background emission was subtracted. The NIRCam images are diffraction limited 
and provide an angular resolution of 0.07$^{\prime\prime}$ (28 au) at 2 {\textmu}m. 
More details on NIRCam data reduction are presented in \cite{HabartE_imaging23}.

\subsubsection*{JWST NIRSpec}

Observations with the NIRSpec \cite{boker2023orbit} integral field unit (IFU, \cite{boker2022near}) 
were obtained as part of the PDRs4All Early Release Science program 1288 \cite{berne2022} on Sep. 10, 2022. 
The observations were processed with version 1.8.2 of the JWST pipeline \cite{jwst_pipeline}. 
We used the F100LP, F170LP, and F290LP filters and the NRSRAPID 
readout pattern, with 5 groups, one integration and 4 dithers, yielding a total integration
time of 257.68s for each filter. 
More details on the observations and data reduction are presented elsewhere \cite{PeetersE_23}. 
We then extracted a spectrum over the full
wavelength range (1 to 5.2 {\textmu}m) in two apertures; one corresponding to
the d203-506 disk, the other to a position near the disk, providing an off-target reference. 
The disk spectrum was obtained with an elliptical aperture
centered on coordinates: right ascension $ {\rm RA} =$5$^{\rm h}$35$^{\rm m}$20$^{\rm s}$.357,
declination ${\rm Dec}=-05^{\circ}25^{\prime}05^{\prime\prime}.81$  (J2000 equinox), 
with dimension length $l=0.52^{\prime\prime}$, height $h=0.38^{\prime\prime}$
and a position angle PA=33~degrees East of North. This aperture is shown in Fig. 2. 
The off-target spectrum was obtained in a circular aperture 
of radius $r$=0.365$^{\prime\prime}$ centered on coordinates: ${\rm RA}=$5$^{\rm h}$35$^{\rm m}$20$^{\rm s}$.370, 
${\rm Dec}=$-5$^{\circ}$25$^{\prime}$04$^{\prime\prime}$.97. 
%We compared the intensities derived in the NIRSpec spectrum 
%to the NIRCam intensities (using the appropriate NIRCam filter 
%transmission curves \cite{NIRCam_filters}) and found an average flux calibration difference 
%of 25\%  between the two instruments, so adopt 25\% as the relative flux uncertainty
%for both instruments.

\subsection*{Radiation field}
\label{sec_rad-field}

d203-506 is situated in the Orion Bar, at an angular distance of 
120$^{\prime\prime}$ from the Trapezium cluster, corresponding to a projected distance 
of 0.25~pc, for a distance to the Orion Nebula of 414~pc \cite{menten2007distance}. 
$\Theta^{\rm 1}$~Ori~C, the most massive star of the Trapezium cluster,
is of spectral type O6 and is the dominant source of UV photons in the nebula. 
Using a synthetic O6 star spectrum from \cite{kurucz1992model}
and applying spherical dilution for a radius equal to the projected distance of 0.25~pc, we calculate an expected
FUV radiation field $G_{\rm 0}=4\times10^{ 4}$ after normalization 
\cite{Habing68}. This calculation adopts a definition \cite{le_petit_model_2006}
of FUV photons as those with energies between 5.17 and 13.60~eV. 
d203-506 is also situated 40$^{\prime\prime}$ west of the B0-2 star $\Theta^{\rm 2}$~Ori~A, 
a projected distance of 0.08~pc. With the same approach, 
this yields $G_{\rm 0}$=8$\times$10$^{\rm 4}$ at the position of d203-506.
These calculations assume that d203-506 and 
the massive stars are located at the same distance from us, 
any offset along the line of sight would reduce the FUV radiation field. We therefore set an upper limit for the radiation field 
received by d203-506 at $G_{\rm 0}$$\leq$1.2$\times$10$^{\rm 5}$.

{  The intensity of the atomic oxygen line at 1.3168 {\textmu}m can also be used to
estimate the radiation field \cite{marconi1998near}.
 This method has been applied to the same NIRSpec observations as we use \cite{PeetersE_23}, which indicated the median radiation field at the position of the ionization front of the Orion Bar has an intensity $G_{\rm 0}$=5.9$\times$10$^{\rm 4}$. The observed average [O~\textsc{i}] intensity at the position d203-506 is 40\% of the value at the ionization front, indicating a radiation field $G_{\rm 0}$=2.4$\times$10$^{\rm 4}$.}
In the rest of the study we adopt $G_{\rm 0}$=2$\times$10$^{\rm 4}$.

\subsection*{Disk dimensions and mass}
\label{sec_derivation-disk-mass}

We derived the disk dimensions from the ALMA continuum emission 
(Fig.~2G), using the \texttt{imfit} task in 
CASA \cite{casa}  with a single Gaussian component. 
This yields a major axis size of 237.6$\pm3.4$~milliarcseconds (mas),
a minor axis size 129.8$\pm3.4$~mas, and a disk position 
angle of 20.3$\pm1.3$~degrees East of North. At a distance of 414~pc \cite{menten2007distance}, 
this implies a physical radius $r_{\rm d}^{\rm ALMA}$=98$\pm1$~au and a disk thickness $E_{\rm d}^{\rm ALMA}$=54$\pm1$~au {\rm for the dust disk}.
We applied the same method to the NIRCam 1.4 {\textmu}m (Fig.~2B)
image, finding radius $234 \pm 80$~mas, 
so $r_{\rm d}^{\rm NIRCam}=97 \pm 13$~au, and thickness $141 \pm 80$~mas, 
that is $E_{\rm d}^{\rm NIRCam}=58 \pm 13$~au.
We also use the HCN ($v=0, J=4 \rightarrow 3$)  maps to derive the gas disk size, finding $r_{\rm g}^{\rm ALMA}$=124~au. 
This value is $\sim 1.26$ times larger than for the dust derived radius, consistent with previous measurements that gas derived radii are $\sim 1.4$ times larger than those 
derived from dust \cite{boyden2020protoplanetary}. 
%Overall the disk size as derived from dust 
%are $r_{\rm d} = 98 \pm 2$~au and $E_{\rm d}= 56 \pm 5$~au.
The central star is not visible, so we derive 
a lower limit for the disk inclination : $i\gtrsim90-\rm{arctan}(\frac{E_{\rm d}}{2r_{\rm d}})$,
finding $i\gtrsim 75^{\circ}$.
Dimensions of d203-506 extracted in this section are summarized in Table~\ref{tab_disk-properties}.
%We therefore adopt $r_{\rm d}=100\pm10$ AU. From the 
%same images, we derive a disk scale height $H=17 \pm 5$ AU. 

{  To derive the total mass of the disk, we adopt a previous method \cite{mann2014alma} so use ALMA 344 GHz continuum observations 
to derive the masses of disks in Orion.
The mass of the disk is:
\begin{equation}
M_{\rm d}=\frac{F_{\rm dust}d^2}{k_{\nu} B_{\nu}(T_{\rm d})},
\label{eq_mass-disk}
\end{equation}
where $F_{\rm dust}$ is the flux of dust emission at 
344~GHz, $d$ is the distance to Orion, 
$k_{\nu}$ is the dust grain opacity at 344~GHz for a gas-to-dust-mass-ratio of 100,
$B_{\nu}$ is the Planck function, and $T_{\rm d}$ 
is the characteristic dust temperature.}
We measure $F_{\rm dust}$= 22.1~mJy from our ALMA 
observations of the dust continuum emission 
(Fig.2G) using the \texttt{imfit} 
task in CASA. Adopting 
$k_{\nu}=0.034$~cm$^2$g$^{-1}$ \cite{beckwith2000dust}, $T_{\rm d}=20$K
 \cite{mann2014alma}, and $d=414$ pc
yields $M_{\rm d}$=11.8~M$_{\rm Jup}$,
where 1 M$_{\rm Jup} = 1.87 \times10^{27}$ kg is the mass of Jupiter. 
This value is consistent with disk masses derived for other disks
in Orion by \cite{mann2014alma}.
{  The derivation of disk masses 
using this approach is highly uncertain. 
First, the adopted value of $k_{\nu}$ is intrinsically 
uncertain; using an alternative opacity of $k_{\nu}=0.058$~cm$^2$g$^{-1}$ from \cite{woitke2016} yields $M_{\rm d}$=6.9~M$_{\rm Jup}$.
Second, Eq.~\ref{eq_mass-disk} is applicable only in the 
case of optically thin emission at 344 GHz, and self absorption
becomes an issue for fluxes above 100 mJy \cite{andrews2005circumstellar}. The flux for d203-506 
is 22 mJy so this effect might be negligible. 
The uncertainty on the mass derivation by this method
has been estimated as $\sim$0.2 dex \cite{andrews2005circumstellar}. The value of $T_d$ is also a source of uncertainty, however the choice of $T_d = 20$ K
has been shown to minimize this \cite{andrews2005circumstellar}. Incorporating the uncertainty associated with 
optical depth of 0.2 dex and on the value of $k_{\nu}$ (using the two values above), we find a range of masses $M_d = 4.4 $ to 18.7~M$_{\rm Jup}$.}

\subsection*{Stellar mass}

{To constrain the stellar mass, we use the gas kinematics as traced by the HCN ($v=0, J=4 \rightarrow 3$) 
emission observed with ALMA. Given the spatial resolution,
and velocity dispersion (expected for a low mass star) we expect this method to provide only an upper limit on the stellar mass.  
Fig.~\ref{Fig_PV-diag} shows a position 
velocity diagram of the HCN ($v=0, J=4 \rightarrow 3$)  emission  and predicted velocities for stellar masses $ M_{\star}=0.1, 0.2, 0.3~{\rm M}_{\odot}$. There is no high velocity emission that would imply $M_{\star} > 0.3~{\rm M}_{\odot}$. 
%We have attempted to use the ``eddy'' software (\url{https://eddy.readthedocs.io/en/latest/}) to 
%determine a precise stellar mass based on the HCN ($J=4 \rightarrow 3$)  rotation 
%map. However, because the SNR is relatively low (at best 10 in the rotation map) 
%and because the scales are small (hence our maps contain a limited
%number of usable pixels), eddy does not converge. 
We used the radiative transfer code \textsc{radmc-3d} version 2.0 to simulate the d203-506 
disk assuming the same values of the stellar mass  
(i.e. 0.1 M$_{\odot}$, 0.2 M$_{\odot}$ and 0.3 M$_{\odot}$). We assumed 
a viscous accretion disk following Keplerian rotation 
using the parameters listed in Table \ref{tab_disk-properties}  (except mass) and a constant HCN abundance fraction of $10^{-8}$ with respect to H. 
The synthetic spectral cubes of the HCN ($v=0, J=4 \rightarrow 3$) line cubes were calculated with a 0.05~km/s 
channel-spacing, then resampled to a 0.1~km/s velocity resolution. 
The spatial axes were convolved with the synthetized 
beam from the ALMA observations.
Fig.~\ref{Fig_rad-transfer} compares the first moment map (velocity centroid) 
of the three synthetic spectral cubes and the observations.
We find that for  a stellar mass $>$ 0.3 M$_{\odot}$, the Keplerian 
rotation velocities are faster than observed, consistent with the upper 
limit provided by the P-V diagram. 
We therefore set an upper limit of $M_{\star} < 0.3~{\rm M}_{\odot}$. }

\subsection*{H$_2$ emitting layer dimensions and surface area}
\label{sec_h2_surface}

{ Fig.~\ref{fig_fit-ellipse} shows the NIRCam F212N image of d203-506, 
which traces emission in the H$_2$ ($v = 1 \rightarrow 2,J=3\rightarrow1$) line. 

The thickness of the H$_2$ emitting surface layer is spatially unresolved. We see 
limb brightening on the contour of the envelope with a thickness close to the beam size. 
With a spatial resolution of 0.07$^{\prime\prime}$, this implies a maximum thickness  
$t_{{\rm H}_2}^{\rm max}=2\times 0.07 \times 414 \approx 60~{\rm au}$ for the H$_2$ emitting layer.

We modelled the observed emission H$_2$ emission ring in Fig.~\ref{fig_fit-ellipse} by 
fitting an ellipse with $r_{{\rm H}_2} = 132$ au and $h_{{\rm H}_2}  = 59$ au. 
For both dimensions, we estimate the uncertainty as one NIRCam pixel, 13 au in physical scale. 
To derive the surface area $S$ of the H$_2$ emitting layer, we a consider spheroid :
\begin{equation}
    S=2 \pi r_{{\rm H}_2}^2 + \frac{\pi h_{{\rm H}_2}^2}{e} \ln{\frac{1+e}{1-e}},
    \label{eq_spheroid}
\end{equation}
where the ellipticity $e=\sqrt{1-h_{{\rm H}_2} ^2/r_{{\rm H}_2}^2}$.  The result is listed in Table S1.}

\subsection*{Gas density and temperature in the PDR}
\label{sec_density-temp-env}

%\subsubsection*{Molecular hydrogen}
\label{sec_h2_PDR}

We detect over 30 ro-vibrational lines of molecular hydrogen in the spectrum of d203-506. We  extracted the intensities 
of these lines by subtracting a linear continuum and fitting Gaussian functions at wavelengths taken from a H$_{\rm 2}$ line list \cite{roueff2019full}. We exclude  lines for which there is an H~\textsc{i} line closer 
than 10$^{-3}~$\textmu m in wavelength, to avoid contamination. We also exclude 
lines with an upper energy level above 2.15 eV to limit contamination 
from shock-excited gas. We only consider a line to be detected if it has an intensity
$>8\times10^{-6}$~erg~cm$^{-2}$~s$^{-1}$~sr$^{-1}$. The measured line intensities are listed in Table~\ref{tab_h2-line-list}. 
To determine the physical conditions of the H$_{\rm 2}$ emitting gas, 
we fit them using the $\textsc{meudon pdr}$ code \cite{le_petit_model_2006}
in a two step process.

First, we use the interstellar medium database (ISMDB) of pre-computed $\textsc{meudon pdr}$ models
(version 1.5.4 \cite{ismdb}),
to determine the PDR parameters that best reproduce the H$_2$ emission lines. 
We assume standard Galactic extinction properties from \cite{Fitzpatrick_Massa1990},
that is a reddening $R_\mathrm{V} = 3.1$, and a ratio of the hydrogen column density ($N_{\rm H}$)
to extinction parameter $E (B-V)$, $N_\mathrm{H} / \mathrm{E}(B-V) = 5.8\times 10^{21}$ cm$^{-2}$.
We consider constant density models with a maximum visual extinction $A_{V}=10$. 
The choice of this parameter has little influence on the results, because most of the H$_{\rm 2}$ emission occurs 
at low $A_{V}$. The free parameters are the FUV radiation field intensity and
the gas density $n_{\rm H}$. Fig.~\ref{fig_h2-meudon} shows the discrepancy measure 
between these models and the observations computed for a range of values for these two parameters. 
The best fitting models have high radiation fields ($G_{\rm 0}>$10$^{\rm 4}$), consistent 
with values derived above via a different method. 
Fig.~\ref{fig_h2-meudon} shows that models 
with densities $n_{\rm H}=1\times$10$^{\rm 5}$~cm$^{-3}$ to $\times$10$^{\rm 7}$~cm$^{-3}$ provide the best fit.

In a second step, we run specific PDR models. 
We use an updated version of the Meudon PDR code, with the addition 
of collisional deexcitation data for H$_2$ in excited vibrational states \cite{Wan2018,Bossion2018}. 
The source code used in this paper is available online \cite{pdrcode}.
We fix the total $A_\mathrm{V} =10$ and intensity of the UV radiation field 
$G_0 = 2\times 10^4$ as derived from the [O~\textsc{i}] line intensity. We run models for 9 
logarithmically-spaced gas density values spanning a smaller range of densities 
($n_{\rm H} \in [10^4,10^8]$~cm$^{-3}$), corresponding to the range favored in the first step above
with a one order of magnitude margin. 

We run these models for two sets of extinction properties and grain size distributions. 
The first set, which we refer to as small grains, uses extinction properties typical 
of dense molecular gas in the Orion Bar \cite{Joblin2018} ($R_\mathrm{V}=5.5$, $N_\mathrm{H} / \mathrm{E}(B-V) = 1.05\times 10^{22}$ cm$^{-2}$, 
the HD 38087 extinction curve \cite{Fitzpatrick_Massa1990}, and a grain size distribution extending from 3 nm to 0.3 {\textmu}m). The second set, which we refer to as large grains, uses larger grain sizes (grain size distribution extending from 20 nm to 1 {\textmu}m, $R_\mathrm{V}=5.9$, $N_\mathrm{H} / \mathrm{E}(B-V) = 1.5\times 10^{22}$ cm$^{-2}$).

We assume all H$_2$ emission lines are emitted in the same layer of the PDR and are all optically thin and account 
for possible geometrical effects (beam dilution of the emitting structure, or inclination of the PDR surface with 
respect to the line-of-sight) with a scaling parameter $\alpha$ (multiplying all observed line intensities) that 
is adjusted simultaneously with density during model fitting. To account for both calibration uncertainties and 
model uncertainties, we consider a multiplicative lognormal error of 50\% on all line intensities and minimize 
the corresponding negative log-likelihood, then convert them to the corresponding reduced $\chi^2$. 

Fig.~\ref{fig_chi2_maps_meudon} shows the $\chi^2$ values of the models fitted to the observations, 
for both extinction settings. We find a bimodal distribution in both cases: one minimum of the 
$\chi^2$ is found at densities $n_\mathrm{H} = 1.5 \times 10^5$ cm$^{-3}$, with $\chi^2=3.0$, 
for the molecular extinction models, and $n_\mathrm{H} = 4.4 \times 10^5$ cm$^{-3}$, with $\chi^2=2.1$, 
for the large grains models), and a second minimum at $n_\mathrm{H} = 5.5 \times 10^6$ cm$^{-3}$, 
with $\chi^2=3.4$, for the molecular extinction models, and $n_\mathrm{H} = 7.2 \times 10^6$ cm$^{-3}$, 
with $\chi^2=3.0$, for the large grains models. Both minima are consistent with the range of values 
determined from the grids discussed above. Fig.~\ref{fig_chi2_maps_meudon} also shows similar $\chi^2$ 
values are found for density values between these two minima (although for lower scaling factor values), 
and for densities up to $\sim10^7$ cm$^{-3}$.

 In these models, we define the emitting layer of the H$_2$ ($v = 1 \rightarrow 2,J=3\rightarrow1$) 
 line as the region in which the line integrated emissivity is 80\% of the total intensity of the line 
 and that has the same emissivity value at its left and right bounds. We measure the width of this 
 emitting layer in each model;  Fig.~\ref{fig_H2_emitting_layer} shows an example. 
 We find that the size constraint  from the observations (that the maximum 
 thickness of the H$_2$ emitting layer is $t_{{\rm H}_2}^{\rm max}=60$, see above) 
 eliminates the lowest density solution for both model sets: it 
 constrains the density to $n_\mathrm{H}>5.5\times10^5$ cm$^{-3}$ for the molecular extinction 
 models, and to $n_\mathrm{H}>9.8\times10^5$ cm$^{-3}$ for the large grains models.

Combining these constraints for the molecular extinction case, our best fitting model has a gas density 
of $n_\mathrm{H} = 5.5 \times 10^6$ cm$^{-3}$ and a scaling factor of 0.8. Its H$_2$ emitting layer is 
2.5 au wide, with an average temperature of $1.24\times 10^3$~K. For the ``large grains'' case, the 
best fitting model has a gas density of $n_\mathrm{H} = 7.2 \times 10^6$ cm$^{-3}$ and a scaling factor 
of 1.4. The H$_2$ emitting layer is 2.8 au wide, with an average temperature of $1.26\times 10^3$ K. 
Fig.~5 shows the H$_2$ line intensities predicted by these two 
best-fitting models compared to the observed intensities. As discussed above, models with similarly 
good $\chi^2$ can be found over a range of density values: density values between the spatial scale 
constraint ($5.5\times10^5$ cm$^{-3}$ for molecular extinction, $9.8\times10^5$ cm$^{-3}$ for large 
grains) and $\sim10^7$ cm$^{-3}$ are found to be compatible with the observations. This implies that 
the overall acceptable range of densities is $n_{\rm H} = 5.5\times 10^{5} $ to $ 1.0\times10^{7}$ cm$^{-3}$.

\subsection*{1D models of external photoevaporation}
\label{sec_model_vs_observations}

To assess the photoevaporation of the d203-506 disk, we compare our derived mass-loss rate to PDR-dynamics calculations 
of external photoevaporative winds using the \textsc{torus-3dpdr} code \cite{Bisbas2015TORUS3DPDR, 2019A&C....27...63H}. 
%This code has previously been used to study external photoevaporation in 1D and 2D models \cite{2016MNRAS.463.3616H, haworth2018fried, 2019MNRAS.485.3895H}. 
The code uses an operator splitting approach in which hydrodynamics and PDR/radiative transfer steps are performed 
iteratively. 
The reduced University of Manchester Institute of Science and Technology (UMIST)
PDR network is used with 33 species and 330 reactions \cite{2013A&A...550A..36M}.
%, chosen to give 
%temperatures with 10\,per cent of the full network \textcolor{red}{Not sure what ``o give 
%temperatures with 10\,per cent of the full network'' means ...}. 
The 1D models the flow structure is solved along a path from the mid-plane to the disk outer edge, then converted into a total mass-loss rate estimate \cite{Adams04Photo}.
We set up models with the parameters expected for the d203-506 disk and external UV field as summarized in Table 
\ref{tab_disk-properties}.  We consider three values of the stellar mass, $M_{\star}$ = 0.1, 0.2, 0.3~M$_{\odot}$ 
and a disk radius of 100 au. Since we do not have constraints on the dust size distribution for d203-506, we vary 
the effective UV absorption cross section of dust at a wavelength $\lambda =0.1$ \textmu m $\sigma_{\rm UV}$ in 
from  10${-23}$ cm$^{-2}$ to 10${-21}$ cm$^{-2}$. $\sigma_{\rm UV}= 10^{-23}$ cm$^{-2}$ corresponds to dust which 
has evolved due to grain growth inside the disk \cite{facchini2016external}, 
while $\sigma_{\rm UV}= 10^{-21}$ cm$^{-2}$ corresponds to small dust grains which have not yet grown as 
found in the interstellar medium in the Orion Nebula \cite{storzer1999}. 
The mass-loss rate from these 1D models as a function $\sigma_{\rm UV}$ 
in the wind is shown in Fig.~\ref{fig:1Dcomparison}. The derived values range between $\dot{M}= 3.3 \times 10^{-8}$ and $ \dot{M}= 1.1 \times 10^{-6}$ M$_\odot$yr$^{-1}$, a range that overlaps the mass-loss rates we derived from the observations.

%where we now
%use a more advanced version of the Meudon code which %includes the latest collisional 

%fix 
%the intensity of the radiation field to 

%provide 
%reasonably good results. We thus adopt $n_{\rm H}$=1-3.5$\times$10$^{\rm 5}$~cm$^{-3}$.
%The gas temperature can be derived from the PDR model. Here, we consider the
%gas temperature at the dissociation front, which is the H$_{\rm 2}$ emitting region,
%and we find $T_{\rm h_{{\rm H}_2} }$=840~K. 

\newpage

 \renewcommand{\thefigure}{{\bf S1}}
\begin{figure}
    \centering
     \includegraphics[width=18cm]{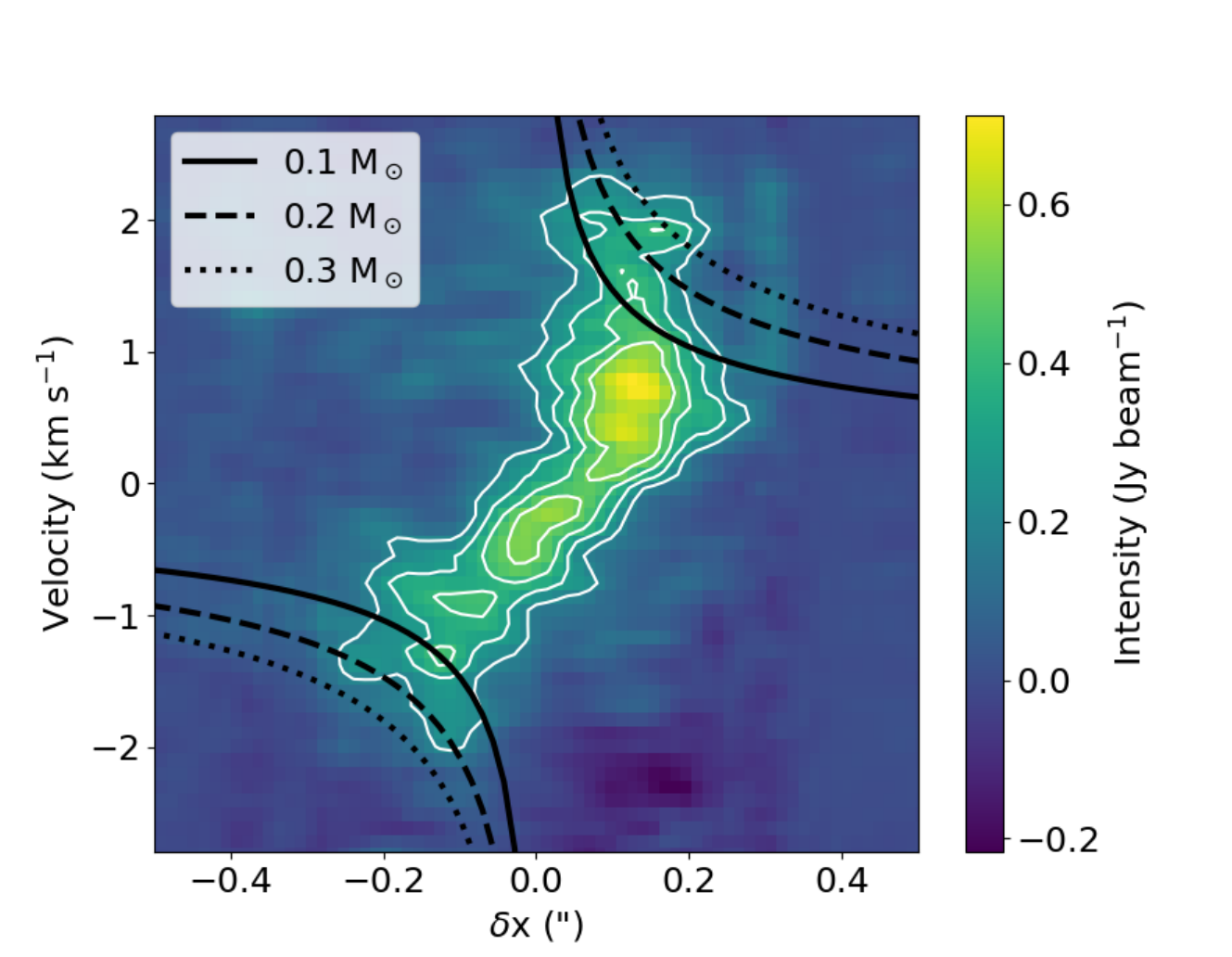}
     \caption{ {\bf Position velocity (PV) diagram of d203-506 along the direction of the disk plane.} $\delta x$ is the angular scale along this axis, with  $\delta x =0$ corresponding to the position of the central star. The black thick curves indicate Keplerian orbital velocities for $M_{\star}=0.1, 0.2, 0.3~{\rm M}_{\odot}$. The white contours correspond to intensities from 0.2 to 0.5 Jy beam$^{-1}$ in steps of 0.1 Jy beam$^{-1}$. 
     Most of the observed emission ends close to the 0.2 M$_{\odot}$ curve.}
     \label{Fig_PV-diag}
\end{figure}

 \renewcommand{\thefigure}{{\bf S2}}
\begin{figure}
    \centering
     \includegraphics[width=18cm]{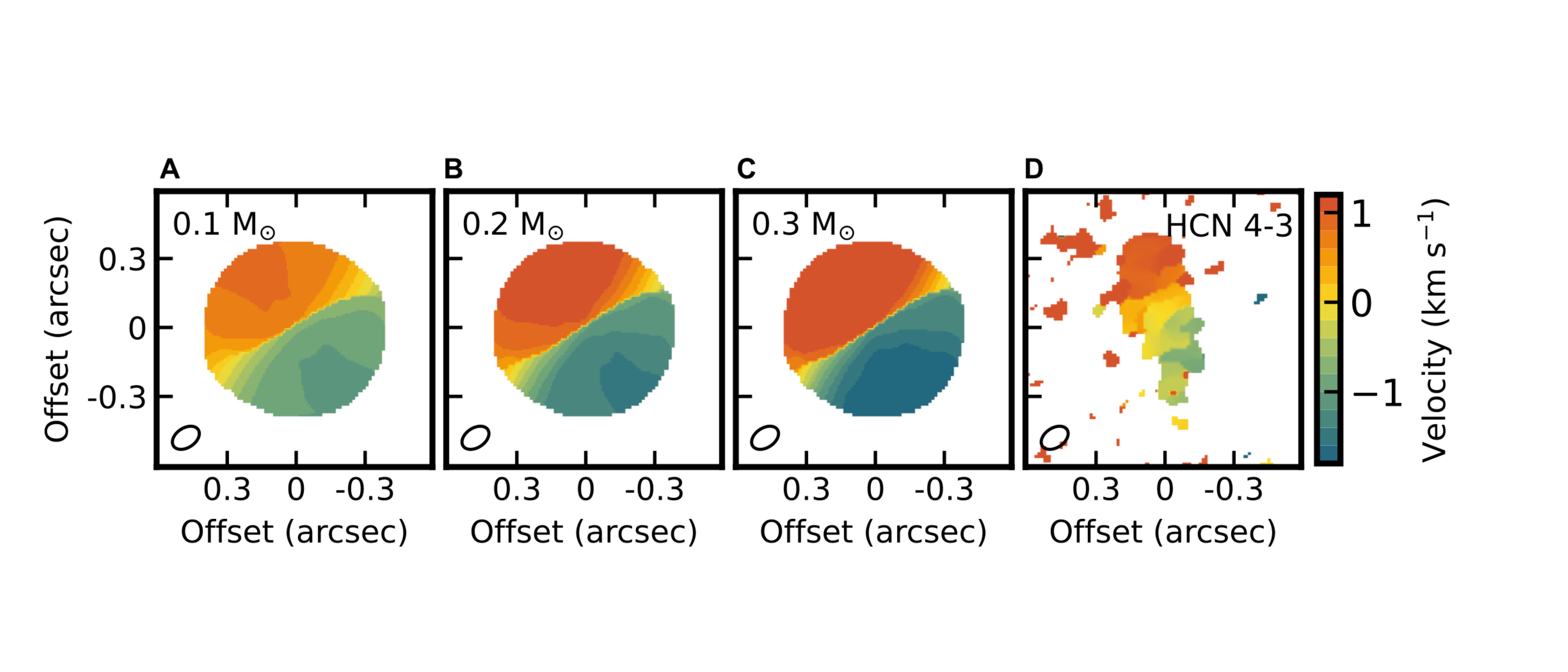}
     \caption{ {\bf Comparison between modelled and observed velocity fields in d203-506}. The panels show the first moment maps from radiative transfer models of the HCN ($v=0, J=4 \rightarrow 3$)  emission of d203-506 assuming $M_{\star}=0.1~{\rm M}_{\odot}$ ({\bf A}), $M_{\star}=0.2~{\rm M}_{\odot}$ ({\bf B}), $M_{\star}=0.3~{\rm M}_{\odot}$ ({\bf C}), compared to ALMA observations ({\bf D}). The offset in the X and Y axis is given in arcseconds with respect to the position of the central star. The ellipse shows the reconstructed beam of the telescope.}
     \label{Fig_rad-transfer}
\end{figure}

\renewcommand{\thefigure}{{\bf S3}}
\begin{figure}[!ht]
    \centering
     \includegraphics[width=10cm]{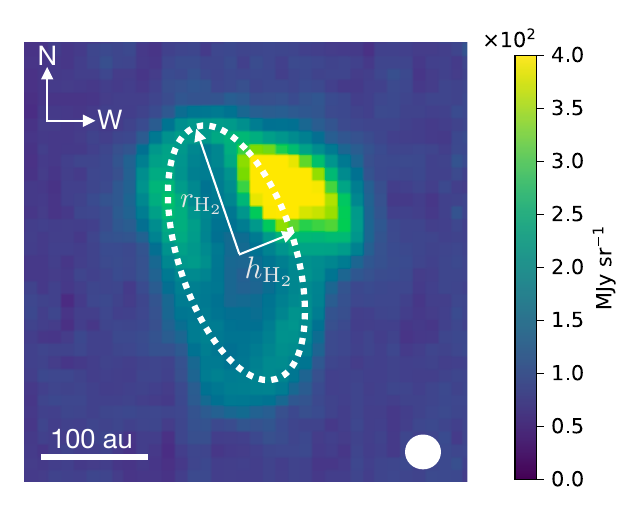}
    \caption{{\bf Image of d203-506 in the NIRCam F212N filter tracing the emission of the H$_2$ ($v = 1 \rightarrow 2,J=3\rightarrow1$) line at 2.12 \textmu m.} The horizontal scale bar is 100 au. The white circle indicates the size of the JWST PSF. The dashed white line shows the elliptical model fitted to the data. The white arrows indicate its major and minor axes. North and West directions are indicated in the upper left corner. The image is centered at coordinates RA =  $5^{\rm h} 35^{\rm m} 20^{\rm s}.357$ and Dec = $-5^{\circ}25^{\prime}05^{\prime\prime}.81$.}
    \label{fig_fit-ellipse}
\end{figure}

 \renewcommand{\thefigure}{{\bf S4}}
\begin{figure}[!ht]
    \centering
     \includegraphics[width=15cm]{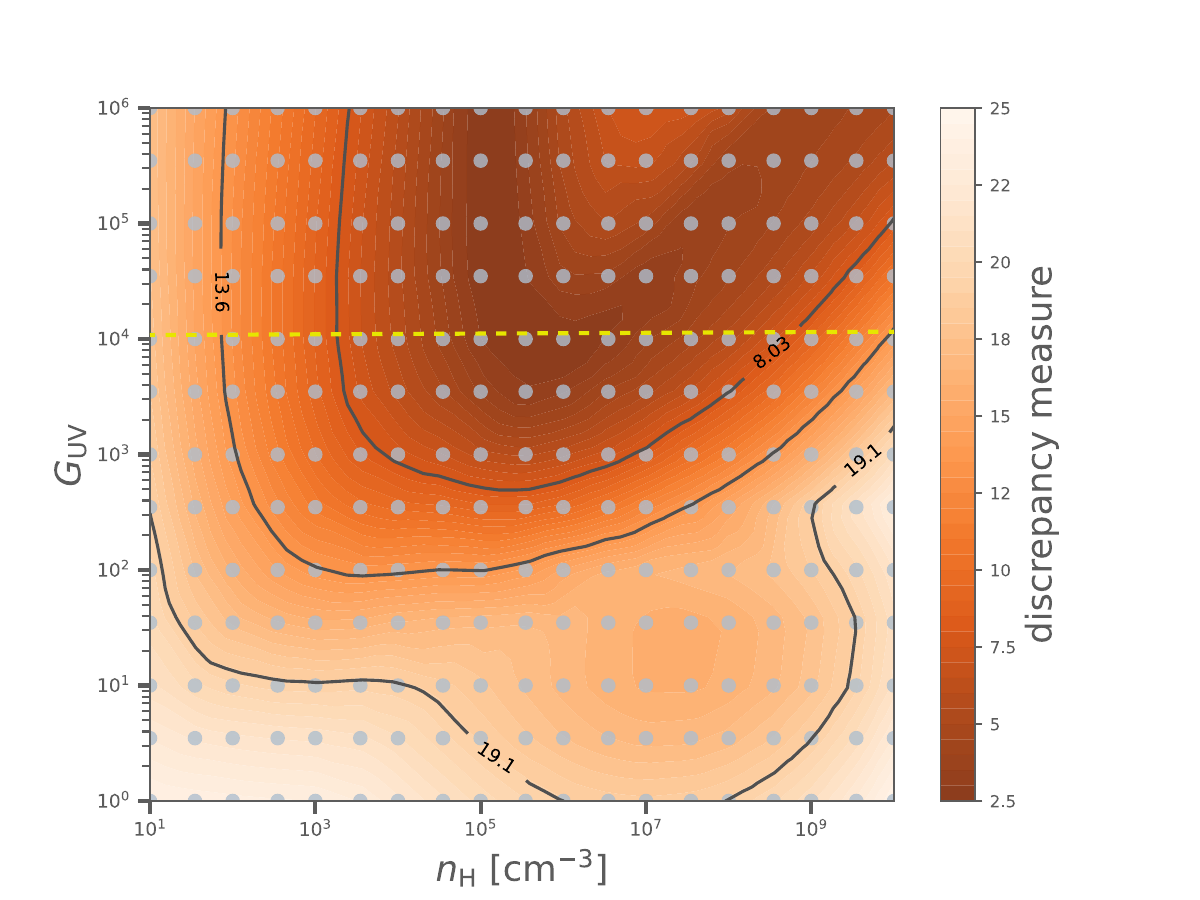}
    \caption{{\bf Results from the PDR model fitting to the H$_{\rm 2}$ ro-vibrational lines.} The map shows a measure of discrepancy (sum of squared distances to the closest bound of the uncertainty intervals for each observed line, in logarithm) for a grid of different UV radiation field strength ($G_{\rm UV}$ in units of the Mathis  ISRF field, that is 1.56 times the Habing field, $G_{\rm UV} = 1.56~G_0$), and gas density ($n_{\rm H}$). Grey circles indicate the models in the grid, while the contours and color map are computed by interpolation. The adopted value for the intensity of the radiation field for d203-506 \cite{methods}, $G_0=2\times10^4 $ corresponding to $ G_{\rm UV} = 1.28\times10^4$ is shown by the yellow horizontal dashed line. 
    For this value of the intensity of FUV radiation, models with densities in the range $n_{\rm H} = 10^{5} - 10^{7}$ cm$^{-3}$ provide the lowest values of the discrepancy measure.}
    %The four best models (accounting for our constraint on the external UV radiation field in d203-506) are circled in yellow.}    %For d203-506, we have $G_{\rm 0}$=4$\times$10$^{\rm 4}$, hence $\chi_{\rm front}\sim$2.6$\times$10$^{\rm 4}$, as indicated by the horizontal dashed line in the plot
    \label{fig_h2-meudon}
\end{figure}

 \renewcommand{\thefigure}{{\bf S5}}
\begin{figure}[!ht]
    \centering
     \includegraphics[width=6.9cm]{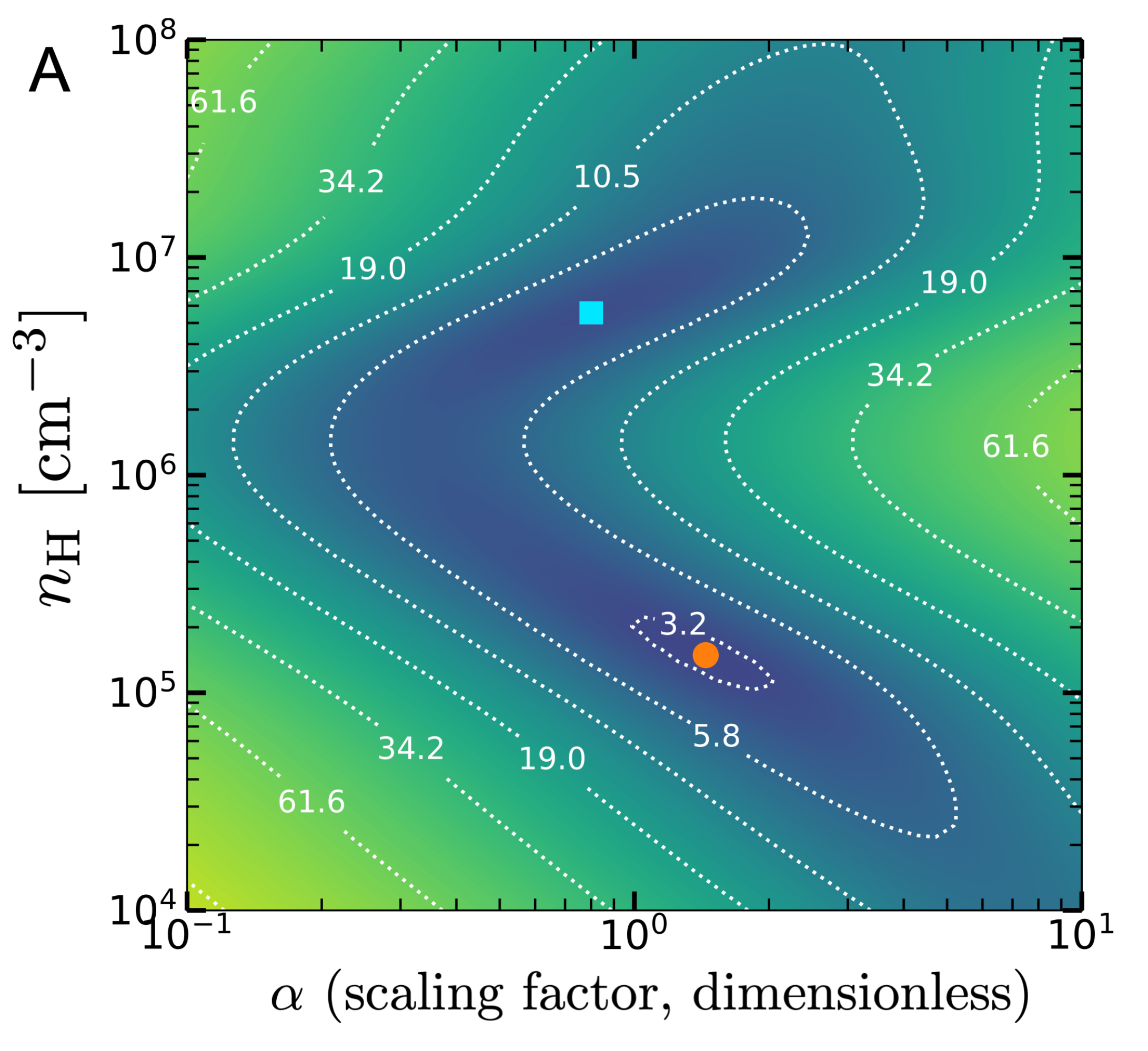}
     \includegraphics[width=8cm]{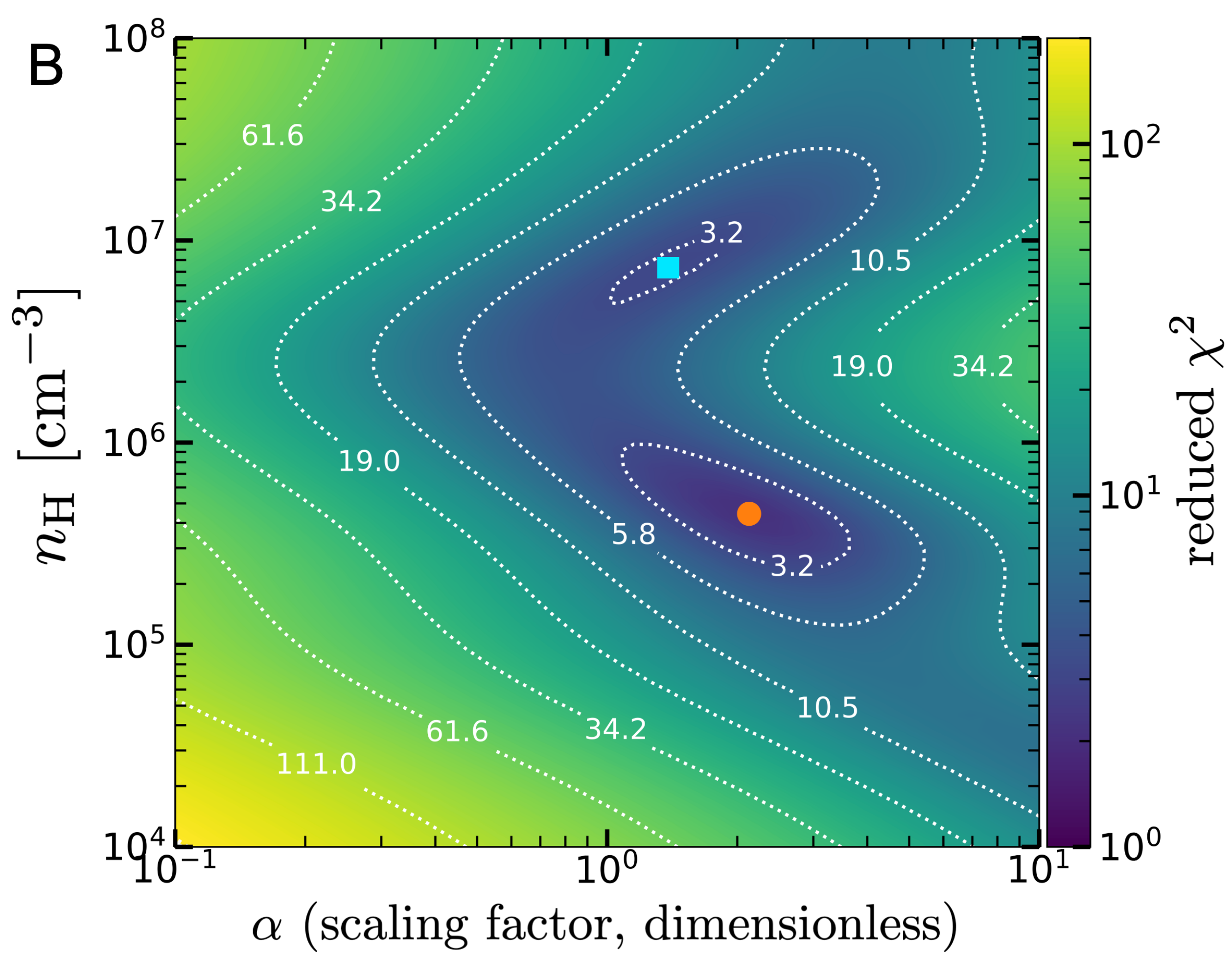}
    \caption{{ \bf  Deviation between observed and modelelled H$_2$ emission intensities. } Reduced $\chi^2$ map, for a grid of values of the gas density $n_\mathrm{H}$ and scaling factor $\alpha$ used in the Meudon PDR models, using small grains ({\bf A}) and large grains ({\bf B}) in the model. Both panels show a clear bimodal distribution, the orange circle indicates the position of minimum in the low density mode and the blue square the position of the minimum of the high density mode. %The orange line shows the isocontour at $\Delta \chi^2 = 1$ above the best $\chi^2$.
    }
    \label{fig_chi2_maps_meudon}
\end{figure}

 \renewcommand{\thefigure}{{\bf S6}}
\begin{figure}[!ht]
    \centering
     \includegraphics[width=10cm]{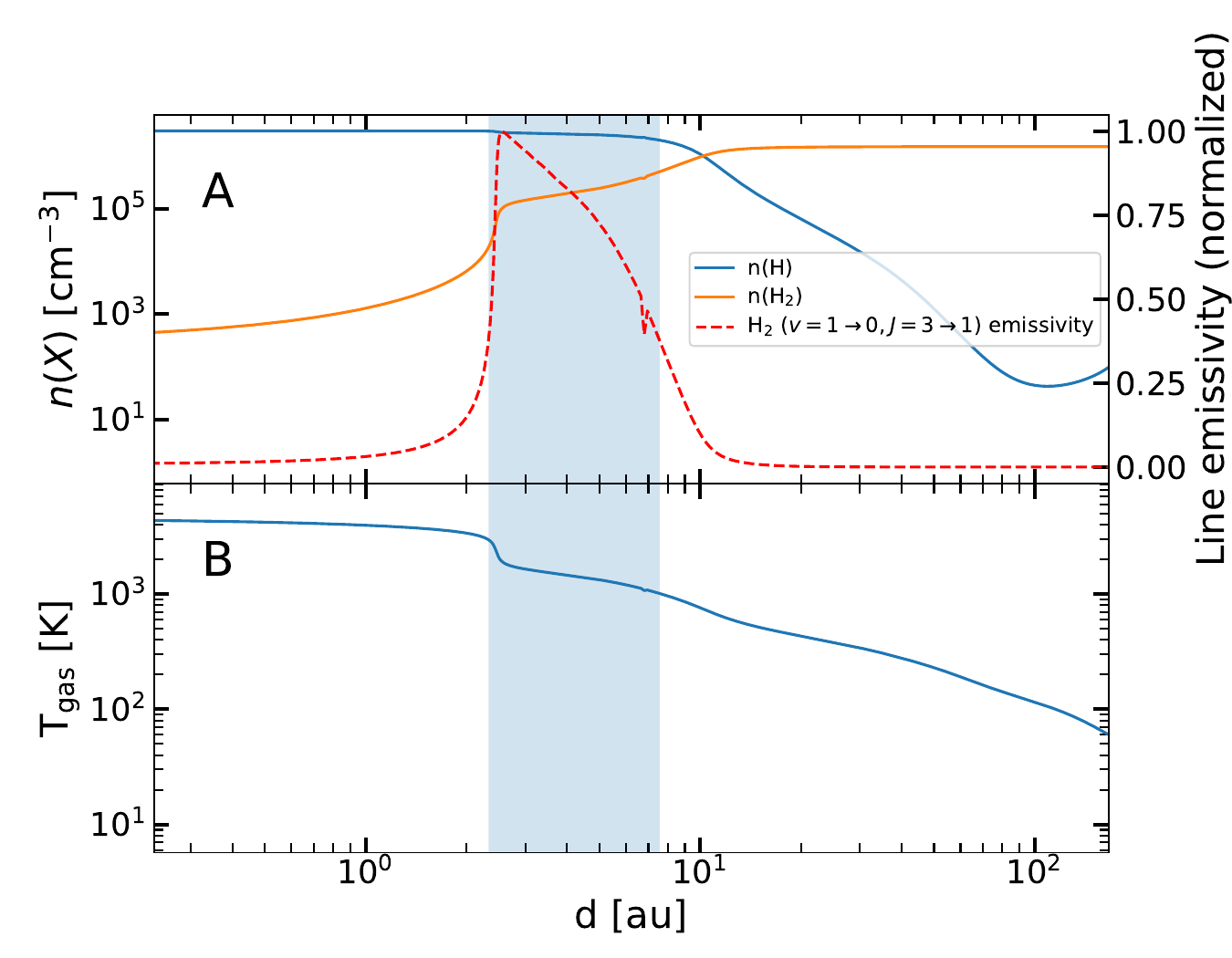}
    \caption{{ {\bf  Spatial structure of a Meudon PDR model (using the small grains setup) for $n_\mathrm{H}=3\times10^6$ cm$^{-3}$}. The UV illumination ($G_0=2\times 10^4$) is from the left side of this plot. Panel ({\bf A}) shows the densities of H (blue) and H$_2$ (orange), both on the left axis. The emissivity of the ($v=1\leftarrow0, J = 3 \leftarrow 1$) line is shown with the red dashed line (right axis). Panel ({\bf B}) shows the gas temperature profile. The blue shaded area shows the H$_2$ emitting layer as defined in the text.}}
    \label{fig_H2_emitting_layer}
\end{figure}

 \renewcommand{\thefigure}{{\bf S7}}
\begin{figure}
    \centering
    \includegraphics[width=\columnwidth]{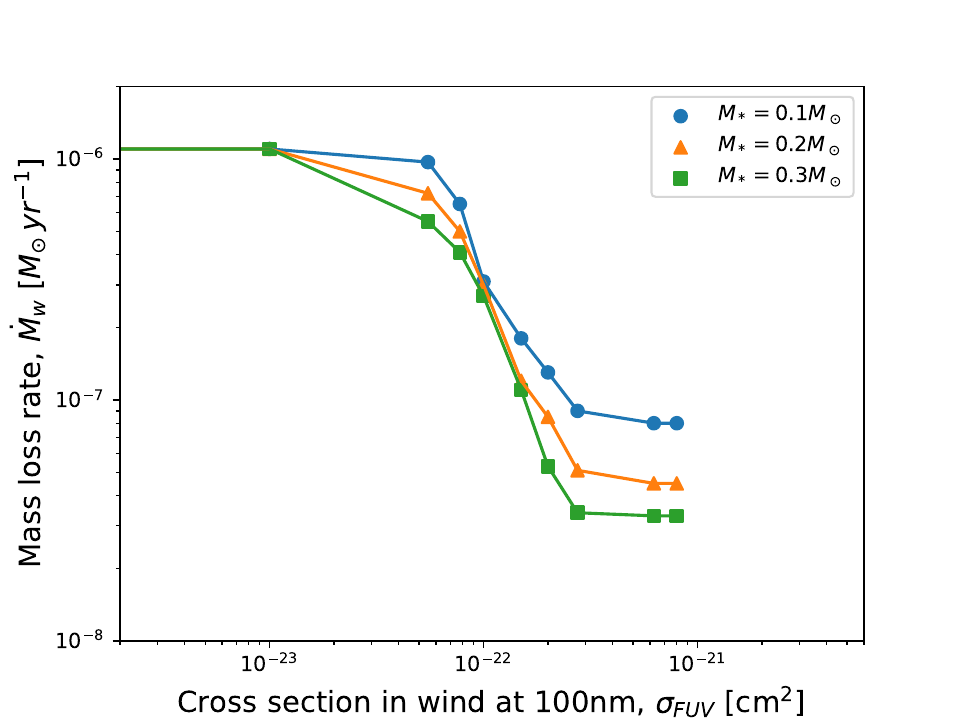}
    \caption{ {\bf Mass-loss rates from 1D \textsc{torus-3dpdr} external photoevaporation calculations as a function of the effective grain cross section in the wind.} The different sets of lines/points correspond to different masses of the central star that the disk orbits.
    }
    \label{fig:1Dcomparison}
\end{figure}

\begin{landscape}
\renewcommand{\thetable}{{  S1}}
\begin{table*}[!h]
\renewcommand{\arraystretch}{1.5} % Default value: 1
\centering
\caption{\textbf{Derived physical properties of d203-506}}
\begin{tabular}{lccc} 
\hline
Parameter & Notation & Value & Method or reference   \\
\hline
%\multicolumn{4}{c}{Disk properties}\\
%\hline
Distance                            & $d$                      & $414\pm7$  pc      & \cite{menten2007distance}   \\
Disk radius (dust emission)         & $r_{\rm d}^{\rm ALMA}$   & $98\pm1$  au &   Dust emission (ALMA)   \\
Disk radius (dust absorption)       & $r_{\rm d}^{\rm NIRCam}$ & $97\pm13$  au&   Dust absorption (NIRCam) \\
Disk radius  (HCN emission)         & $r_{\rm g}^{\rm ALMA}$   & $124\pm3$  au&  HCN emission (ALMA)       \\
Disk thickness (dust emission)      & $E_{\rm d}^{\rm ALMA}$   & $54\pm1$ au  & Dust emission (ALMA)    \\
Disk thickness (dust absorption)    & $E_{\rm d}^{\rm NIRCam}$ & $58\pm13$ au & Dust absorption (NIRCam)    \\
%Disk thickness   (gas)               & $E_{\rm g}^{\rm ALMA}$               & $82\pm5$ au   &  HCN (4-3) emission (ALMA)    \\
Radius of H$_2$ emitting layer      & $r_{{\rm H}_2}$         & $127 \pm 13$ au       & H$_2$ ($v=2\rightarrow1, J=3\rightarrow1$) (NIRCam)     \\
Height of H$_2$ emitting layer      & $h_{{\rm H}_2}$         & $56 \pm 13$ au        & H$_2$ ($v=2\rightarrow1, J=3\rightarrow1$) (NIRCam)     \\
Max. thickness of the H$_2$ emitting layer & $t_{{\rm H}_2}^{\rm max}$ & $ 60 \pm 13$ au & H$_2$ ($v=2\rightarrow1, J=3\rightarrow1$)\\
Disk inclination                    & $i$             &  $\gtrsim 75^{\circ}$ & Dust absorption (NIRCam) \\
Disk mass                           & $M_{\rm d}$     &  $4.4 $ to $ 18.7$ Jupiter mass (M$_{\rm Jup}$) & Dust emission (ALMA)  \\
Stellar mass                        & $M_{\star}$     &  $ < 0.3 $ M$_{\odot}$& HCN ($J=4 \rightarrow 3$)  emission  (ALMA) \\
Ambient radiation field             & $G_0$           & $2.4 \times10^4$      & [O~\textsc{i}] line emission (NIRSpec) \\
Gas temperature (PDR)               & $T_{\rm gas}$   &  1240 to 1260~K    & H$_2$ rovib. lines (NIRSpec)    \\
Gas density    (PDR)                & $n_{\rm H}$     &  $5.5\times10^{5} $ to $ 1.0\times10^{7}$ cm$^{-3}$ & H$_2$ rovib. lines  (NIRSpec)\\
Surface of H$_2$ emitting layer     & $S$             & $1.3 \pm 0.4 \times 10^{5}$ au$^2$ & H$_2$ ($v=2\rightarrow1, J=3\rightarrow1$)(NIRCam) \\ 
Mass-loss rate                      & $\dot{M}$&    $1.4\times 10^{-7}$ to $4.6\times 10^{-6}$ M$_{\odot}$/yr & \cite{methods} \\
%Depletion timescale             & $\tau$       & $xx - xx $ Myr & -\\
\hline         
\end{tabular}
\label{tab_disk-properties}
\end{table*} 
\end{landscape}

\renewcommand{\thetable}{{  S2}}

\begin{table}[h!]
\centering
\caption{\textbf{Detected H$_{\rm 2}$ lines towards the PDR in d203-506.} Rest wavelengths are from \cite{roueff2019full}.}
\vspace{-0.2cm}
\begin{tabular}{llccc} 
\hline
Line              & Transition & Rest wavelength  & Intensity                                 & Uncertainty (1$\sigma$)\\
(quantum levels)  & (abreviation)  & ($\rm{\mu}$m)    & (erg~cm$^{\rm -2}$~s$^{-1}$~sr$^{-1}$) & (erg~cm$^{\rm -2}$~s$^{-1}$~sr$^{-1}$) \\
\hline
($v=0, J=10\rightarrow8$)             & 0-0 S(8)   & 5.0531 & 9.24$\times 10^{-5}$ & 1.19$\times 10^{-7}$ \\
($v=0, J=11\rightarrow9$)             & 0-0 S(9)   & 4.6946 & 1.51$\times 10^{-4}$ & 4.59$\times 10^{-8}$ \\
($v=0, J=12\rightarrow10$)            & 0-0 S(10)  & 4.4097 & 2.73$\times 10^{-5}$ & 1.63$\times 10^{-7}$ \\
($v=0, J=13\rightarrow11$)            & 0-0 S(11)  & 4.1810 & 3.83$\times 10^{-5}$ & 3.49$\times 10^{-8}$ \\
($v=0, J=15\rightarrow13$)            & 0-0 S(13)  & 3.8461 & 1.88$\times 10^{-5}$ & 2.05$\times 10^{-8}$ \\
($v=0, J=17\rightarrow15$)            & 0-0 S(15)  & 3.6261 & 8.66$\times 10^{-6}$ & 2.69$\times 10^{-8}$ \\
($v=1\rightarrow0, J=1\rightarrow3$)  & 1-0 O(3)   & 2.8025 & 1.28$\times 10^{-3}$ & 1.16$\times 10^{-6}$ \\
($v=1\rightarrow0, J=2\rightarrow4$)  & 1-0 O(4)   & 3.0038 & 2.00$\times 10^{-4}$ & 3.63$\times 10^{-8}$ \\
($v=1\rightarrow0, J=2\rightarrow0$)  & 1-0 S(0)   & 2.2232 & 3.21$\times 10^{-4}$ & 1.32$\times 10^{-6}$ \\
($v=1\rightarrow0, J=3\rightarrow5$)  & 1-0 O(5)   & 3.2349 & 3.58$\times 10^{-4}$ & 1.57$\times 10^{-7}$ \\
($v=1\rightarrow0, J=3\rightarrow1$)  & 1-0 S(1)   & 2.1218 & 1.06$\times 10^{-3}$ & 1.87$\times 10^{-6}$ \\
($v=1\rightarrow0, J=4\rightarrow6$)  & 1-0 O(6)   & 3.5008 & 3.61$\times 10^{-5}$ & 7.48$\times 10^{-9}$ \\
($v=1\rightarrow0, J=4\rightarrow2$)  & 1-0 S(2)   & 2.0337 & 2.84$\times 10^{-4}$ & 1.75$\times 10^{-6}$ \\
($v=1\rightarrow0, J=5\rightarrow7$)  & 1-0 O(7)   & 3.8074 & 7.47$\times 10^{-5}$ & 3.89$\times 10^{-9}$ \\
($v=1\rightarrow0, J=5$)              & 1-0 Q(5)   & 2.4547 & 1.78$\times 10^{-4}$ & 1.47$\times 10^{-6}$ \\
($v=1\rightarrow0, J=5\rightarrow3$)  & 1-0 S(3)   & 1.9575 & 5.69$\times 10^{-4}$ & 2.79$\times 10^{-6}$ \\
($v=1\rightarrow0, J=6$)              & 1-0 Q(6)   & 2.4755 & 5.16$\times 10^{-5}$ & 6.69$\times 10^{-7}$ \\
($v=1\rightarrow0, J=6\rightarrow4$)  & 1-0 S(4)   & 1.8919 & 7.21$\times 10^{-5}$ & 1.16$\times 10^{-6}$ \\
($v=1\rightarrow0, J=7\rightarrow5$)  & 1-0 S(5)   & 1.8357 & 1.56$\times 10^{-4}$ & 3.15$\times 10^{-6}$ \\
($v=1\rightarrow0, J=8$)              & 1-0 Q(8)   & 2.5280 & 1.40$\times 10^{-5}$ & 1.02$\times 10^{-6}$ \\
($v=1\rightarrow0, J=8\rightarrow6$)  & 1-0 S(6)   & 1.7880 & 2.78$\times 10^{-5}$ & 3.29$\times 10^{-6}$ \\
($v=1\rightarrow0, J=9\rightarrow7$)  & 1-0 S(7)   & 1.7479 & 1.18$\times 10^{-5}$ & 5.68$\times 10^{-7}$ \\
($v=1\rightarrow0, J=12\rightarrow10$)& 1-0 S(10)  & 1.6664 & 1.93$\times 10^{-5}$ & 2.17$\times 10^{-6}$ \\
($v=2\rightarrow0, J=0\rightarrow2$)  & 2-0 O(2)   & 1.2932 & 1.97$\times 10^{-5}$ & 3.05$\times 10^{-6}$ \\
($v=2\rightarrow1, J=0\rightarrow2$)  & 2-1 O(2)   & 2.7861 & 2.77$\times 10^{-5}$ & 5.09$\times 10^{-7}$ \\
($v=2\rightarrow0, J=1\rightarrow3$)  & 2-0 O(3)   & 1.3354 & 2.21$\times 10^{-5}$ & 3.17$\times 10^{-6}$ \\
($v=2\rightarrow0, J=1$)              & 2-0 Q(1)   & 1.2383 & 2.74$\times 10^{-5}$ & 1.50$\times 10^{-6}$ \\
($v=2\rightarrow1, J=1\rightarrow3$)  & 2-1 O(3)   & 2.9740 & 2.84$\times 10^{-5}$ & 1.98$\times 10^{-7}$ \\
($v=2\rightarrow1, J=1$)              & 2-1 Q(1)   & 2.5509 & 4.39$\times 10^{-5}$ & 6.45$\times 10^{-7}$ \\
($v=2\rightarrow0, J=2\rightarrow0$)  & 2-0 S(0)   & 1.1895 & 1.47$\times 10^{-5}$ & 3.12$\times 10^{-6}$ \\
($v=2\rightarrow1, J=2\rightarrow4$)  & 2-1 O(4)   & 3.1898 & 8.57$\times 10^{-6}$ & 1.45$\times 10^{-7}$ \\
($v=2\rightarrow1, J=3\rightarrow5$)  & 2-1 O(5)   & 3.4378 & 9.89$\times 10^{-6}$ & 2.31$\times 10^{-8}$ \\
($v=2\rightarrow1, J=3$)              & 2-1 Q(3)   & 2.5698 & 1.41$\times 10^{-5}$ & 5.04$\times 10^{-7}$ \\
($v=2\rightarrow1, J=3\rightarrow1$)  & 2-1 S(1)   & 2.2477 & 2.59$\times 10^{-5}$ & 7.90$\times 10^{-7}$ \\
($v=2\rightarrow1, J=5\rightarrow3$)  & 2-1 S(3)   & 2.0734 & 2.16$\times 10^{-5}$ & 1.97$\times 10^{-6}$ \\
($v=2\rightarrow0, J=7\rightarrow5$)  & 2-0 S(5)   & 1.0851 & 1.19$\times 10^{-5}$ & 1.87$\times 10^{-6}$ \\
($v=2\rightarrow0, J=8\rightarrow6$)  & 2-0 S(6)   & 1.0732 & 2.20$\times 10^{-5}$ & 2.59$\times 10^{-6}$ \\
($v=2\rightarrow1, J=9\rightarrow7$)  & 2-1 S(7)   & 1.8528 & 2.02$\times 10^{-5}$ & 2.89$\times 10^{-6}$ \\
($v=3\rightarrow2, J=3\rightarrow5$)  & 3-2 Q(5)   & 2.7692 & 9.40$\times 10^{-6}$ & 3.25$\times 10^{-7}$ \\
\hline         
\end{tabular}
\label{tab_h2-line-list}
\end{table}

\end{document}